\documentclass[a4paper,onecolumn]{revtex4}
\usepackage{bm, multirow}
\usepackage{amsmath,amssymb, mathrsfs}
\usepackage{psfrag, color, graphicx}
\usepackage{subfigure}
\newcommand{\etal}{\emph{et al.}}

\newcommand{\Tr}{\mathrm{Tr}\;}

\newcommand{\Utwo}{\ensuremath{U(2)_L \times U(2)_R}}
\newcommand{\HF}{Hartree--Fock}

\begin{document}

\title{Chiral and U(1) axial symmetry restoration in 
  linear sigma models with two quark flavors}
\author{Stefan Michalski}
\email{stefan.michalski@uni-dortmund.de}
\affiliation{Institut f\"ur Physik, Universit\"at Dortmund,
      D-44221 Dortmund, Germany}
\preprint{DO-TH 06/01}
\date{April 3, 2006}

\begin{abstract}
  We study the restoration of chiral symmetry in
  linear sigma models with two quark flavors. The models taken
  into consideration have a \Utwo\ and an
  $O(4)$ internal symmetry. The physical mesons of these models
  are $\sigma$, pion, $\eta$ and $a_0$ where the latter
  two are not present in the $O(4)$ model.
  Including two-loop contributions through sunset graphs
  we calculate the 
  temperature behavior of the order parameter and the
  masses for explicit chiral symmetry breaking and in the
  chiral limit.
  Decay threshold effects
  introduced by the sunset graphs alter the temperature
  dependence of the condensate and consequently that
  of the masses as well. 
  This correctly reproduces a second-order phase
  transition for the $O(4)$ model and for the
  \Utwo\ model with an axial $U(1)$ anomaly as expected from
  universality class arguments.
  Chiral symmetry tends to be restored
  at higher temperatures in the two-loop approximation
  than in the \HF\ approximation.
  To model a restoration of the axial $U(1)$ symmetry
  we imply a temperature-dependent anomaly parameter that
  sharply drops at about 175~MeV. This
  triggers the restoration of chiral symmetry before the
  full symmetry is restored and lowers the transition
  temperatures significantly below 200~MeV.
  
\end{abstract}
\maketitle

\section{Introduction}
\label{sec:intro}
The Lagrangian of massless quantum chromodynamics (QCD) 
with $N_f$ quark flavors
has a chiral $SU(N_f)_L \times SU(N_f)_R \times U(1)_A \times U(1)_V$ 
symmetry. 
A chiral quark condensate 
$\langle \bar{q} q\rangle \approx (300\,\mathrm{MeV})^3$ spontaneously
breaks the $SU(N_f)_A\times U(1)_A \simeq U(N_f)_A$ part of the symmetry
and generates $N_f^2$ Goldstone bosons. Apart from that there is also 
a violation of the $U(1)_A$ symmetry by instantons
\cite{'tHooft:1976up,'tHooft:1976fv,Pisarski:1983ms} giving
mass to one of the Goldstone bosons. The $U(1)_V$ (vector) symmetry 
represents baryon
number conservation, is always fulfilled and therefore will not
be considered here. 
Adding mass terms like $m_q\, \bar{q} q$
to the QCD Lagrangian breaks the symmetry explicitly
and gives all Goldstone bosons a mass, making them
pseudo-Goldstone bosons.

At high temperatures chiral symmetry is expected to be restored. 
For finite quark masses this happens in a crossover transition
such that the symmetry is (almost) restored when the temperature
(to the third power) is of the order of the condensate
$\langle\bar{q}q\rangle^{1/3} \approx 300\,\mathrm{MeV}$.
Recently, lattice QCD has been able to determine the critical
temperature of the chiral phase transition. For three flavors 
it has been found to be in the
vicinity of 155~MeV while for two flavors it is about 170~MeV
for a vanishing quark chemical 
potential~\cite{Karsch:2001cy,Laermann:2003cv}.
In spite of being a challenging first-principle approach to
QCD lattice calculation suffer from technical difficulties
for small quark masses~\cite{Fodor:2004nz} or for
a chemical potential of the order of the temperature
or larger~\cite{deForcrand:2002ci}.

A different nonperturbative approach to QCD is the construction
of low-energy effective theories of hadrons with the same chiral 
$U(N_f) \times U(N_f)$ symmetry.
The color degrees of freedom are integrated out so that
the low-energy behavior of QCD is governed by the lightest hadrons
which are scalar and pseudoscalar mesons with, in general, 
light quark content. These particles can be found in
linear sigma models~\cite{Gell-Mann:1960np}.
Since these models have the same symmetry as the 
underlying fundamental theory and thus belong to the same
universality class they can be used to study
the dynamics of phase transitions at finite temperature. 
Pisarski and Wilczek~\cite{Pisarski:1983ms} found that
there can be a second-order phase transition in presence
of an explicitly broken $U(1)_A$ symmetry, whereas without this
axial $U(1)$ anomaly the transition is of first order.
The two-loop approximation investigated within this
article will correctly reproduce these features in both the \Utwo\ and
the $O(4)$ model.

Linear sigma
models cannot be solved analytically so one has to make use
of approximations. One problem arising at finite temperature
is the breakdown of perturbation theory;
at a temperature $T$, a (perturbative) expansion
in powers of a coupling $g$ yields a new mass scale $gT$ that
occurs in the denominators of loop graphs and cancels powers
of the coupling constant in the perturbation expansion
\cite{Dolan:1973qd,Braaten:1989kk,Braaten:1989mz,Parwani:1991gq}. 
So, terms
of all orders of the coupling
must be taken into account via resummation 
to avoid these unwanted cancellations.
The resummation scheme we apply here is the so-called
\emph{two-particle point-irreducible} (2PPI) effective action
introduced by Verschelde and Coppens~\cite{Verschelde:1992bs}. 
Up to the level of the
\HF\ approximation it is identical to the
\emph{two-particle irreducible} (2PI) effective action formalism by Cornwall,
Jackiw and Tomboulis~\cite{Cornwall:1974vz}.

Linear sigma models with a $U(N_f)\times U(N_f)$ symmetry
and two to four quark flavors 
have been studied in the
\HF\ or Hartree approximation within the last more than 25
years \cite{Roder:2003uz,Lenaghan:2000ey,Schaffner-Bielich:1999uj,
  Geddes:1979nd}. 
The $O(N)$ model has received even greater attention; it has
been analyzed using different resummation techniques, where various
authors used local resummations~\cite{Lenaghan:1999si,Dolan:1973qd,
  Baacke:2002pi,Petropoulos:1998gt,Patkos:2002xb,Chiku:1998kd,Nemoto:1999qf,
  Smet:2001un,Verschelde:2000ta}, while, nowadays,
nonlocal schemes, like the two-particle irreducible effective action
\cite{Cornwall:1974vz}, have become popular as 
well~\cite{Parwani:1991gq,deGodoyCaldas:2001mb,Roder:2005vt,
  Roder:2005qy,Baacke:2004xm,Baacke:2004dp,
  Andersen:2004ae}.
Furthermore, renormalization has become a heavily studied issue in 
this context~\cite{vanHees:2001ik,VanHees:2001pf,vanHees:2002bv,
  Berges:2004hn,Blaizot:2003an}. 

The \HF\ approximation as well as the
two-loop approximation of the 2PPI effective 
action
violate Goldstone's theorem because the formalism's variational
parameter associated to the Goldstone boson mass achieves
finite values at temperatures above 
zero~\cite{Baacke:2002pi,Verschelde:2000ta,Nemoto:1999qf}.
This problem can be overcome either by looking at the external 
(or physical) propagators~\cite{Baacke:2002pi,vanHees:2002bv}
--- the derivatives of the one-particle irreducible
(1PI) effective action --- or 
by a construction described by Ivanov \etal~\cite{Ivanov:2005bv}. 
For a
renormalization group invariant approach to the $O(N)$ model
see recent work of
Destri and Sartirana~\cite{Destri:2005se,Destri:2005qm}.

The $U(N_f)_L \times U(N_f)_R$ linear sigma model contains
two $U(N_f)$ isospin multiplets --- a scalar and a pseudoscalar one ---
 each of which is decomposed
into an isosinglet and an $(N_f^2-1)$-dimensional isospin multiplet. For
two flavors and 
unbroken isospin symmetry ($m_u=m_d$) we obtain four
different mesons in the model, $\sigma$ 
[called $f_0(600)$ nowadays] with an isotriplet of (identical) $a_0$ bosons
in the scalar sector, and $\eta$ with three pions
in the pseudoscalar sector. 
The $O(N)$ linear sigma model only consists of a $\sigma$
meson and $N-1$ pions and is, for $N=N_f^2=4$, a limiting case of the
\Utwo\ model for an infinitely strong $U(1)_A$ anomaly. 

This article is organized as follows. In 
Secs.~\ref{sec:utwo-linear-sigma} and~\ref{sec:on-linear-sigma}
 we describe the
\Utwo\ and the $O(N)$ linear sigma model and their pattern of symmetry 
breaking. Section~\ref{sec:numerical-results} deals with 
parameter fixing and numerical results
in both models. Finally, in Sec.~\ref{sec:conclusions-outlook}
 we draw our conclusions and give an outlook.
There is also an Appendix in which more details about the computation
of the effective action of the \Utwo\ model are given.


\section{The \Utwo\ linear sigma model}
\label{sec:utwo-linear-sigma}

\subsection{Classical action}

The Lagrangian of the $U(N_f)_R \times U(N_f)_L$ linear sigma model
is given by
\begin{equation}
  \label{eq:langrange1}
  \begin{split}
    \mathscr{L}[\Phi] =\ & \Tr ( \partial_\mu \Phi^\dag \partial^\mu \Phi
    - m^2 \Phi^\dag \Phi) - \lambda_1 [ \Tr (\Phi^\dag \Phi)]^2 
    - \lambda_2 \Tr[ (\Phi^\dag \Phi)^2]\\
    &+ c [\det \Phi + \det \Phi^\dag]
    + \Tr [H (\Phi+\Phi^\dag)]\ .
  \end{split}
\end{equation}
The field $\Phi$ is a complex $N_f\times N_f$ matrix containing the scalar and
pseudoscalar mesons,
\begin{equation}
  \label{eq:phidef}
  \Phi = T_a (\sigma_a + i \pi_a).
\end{equation}
Here $\sigma_a$ are the scalar fields with $J^P=0^+$ while $\pi_a$ denotes
the pseudoscalar ones with $J^P=0^-$. 
The last term in the Lagrangian~(\ref{eq:langrange1}) 
describes the interaction with an external field $H$ that breaks the symmetry
explicitly, 
\begin{equation}
  \label{eq:Hdef}
  H = T_a h_a\ .
\end{equation}
$T_a$ are the generators of the group
$U(N_f)$ such that $\Tr (T_a T_b) = \delta_{ab}/2$. The $U(N_f)$ algebra
is fulfilled 
\begin{subequations}
  \label{eq:U2algebra}
  \begin{eqnarray}
    [T_a,T_b] &=& i f_{abc}\ T_c \\
    \{T_a,T_b\} &=& d_{abc}\ T_c
  \end{eqnarray}
  where $f_{abc}$ and $d_{abc}$ are the antisymmetric and symmetric
  structure constants of $U(N_f)$ and $a,b,c=0,\dots,N_f^2-1$. 
  They are identical to those of $SU(N_f)$ (with all indices starting
  from one), however for $U(N_f)$ there is in addition
  \begin{equation}
    f_{ab0} = 0\ , \qquad d_{ab0} = \sqrt{\frac{2}{N_f}}\ \delta_{ab}
    \ .
  \end{equation}
\end{subequations}

In the following we will deal only with the case $N_f=2$ which reduces the
structure constants to 
\begin{equation}
  \label{eq:StrukturkonstantenU2}
  f_{ijk} = \varepsilon_{ijk} \quad\text{and}\quad d_{ijk}=0
  \quad\text{for}\quad i,j,k \in \{1,2,3\}
  \ ,
\end{equation}
where $\varepsilon_{ijk}$ is the Levi--Civita symbol.
The usual identification of the physical bosons for $N_f=2$ is
(see, \textit{e.g.}, Ref.~\cite{Roder:2003uz}) 
\begin{equation}
  \label{eq:physical_bosons}
  \Phi
  = \frac{1}{\sqrt{2}}\left(
    \begin{array}{cc}
      \frac{1}{\sqrt{2}} \left( \sigma + a_0^0 \right) & a_0^+ \\
      a_0^- &  \frac{1}{\sqrt{2}} \left( \sigma - a_0^0 \right)
    \end{array}
  \right)
  + 
  \frac{i}{\sqrt{2}}\left(
    \begin{array}{cc}
      \frac{1}{\sqrt{2}} \left( \eta + \pi^0 \right) & \pi^+ \\
      \pi^- &  \frac{1}{\sqrt{2}} \left( \eta - \pi^0 \right)
    \end{array}
  \right)
  \ .
\end{equation}
Since isospin symmetry is left untouched the masses of all particles
of one isovector are identical, \emph{i.e.}, $m_{a_0^0}=m_{a_0^\pm}$
and $m_{\pi^0}=m_{\pi^\pm}$.

\subsection{Breaking the symmetry}
The first three terms of the Lagrangian~(\ref{eq:langrange1}) are
invariant under the group $U(2)_L\times U(2)_R \simeq 
SU(2)_V\times SU(2)_A \times U(1)_A \times U(1)_V$. 
The $U(1)_V$ vector symmetry reflects baryon number conservation of
QCD. We will not deal with this symmetry in this paper as it is always
conserved.
Chiral symmetry is spontaneously broken if the vacuum
expectation value of the field $\Phi$ does not vanish
\begin{equation}
  \langle \Phi \rangle = T_a \phi_a\ .
\end{equation}
The vacuum should be of even parity, so only 
$\phi_a=\langle \sigma_a \rangle$ is allowed.
According to a theorem by Vafa and Witten~\cite{Vafa:1983tf}
global vector-like symmetries (isospin, baryon number)
cannot be broken spontaneously. So, the remaining
symmetry must be, at least,  $SU(2)_V$. 
Spontaneously breaking $SU(2)_A \times U(1)_A$ 
yields four Goldstone bosons, $\eta$ and three pions.
The determinants in the Lagrangian~(\ref{eq:langrange1}) break the
$U(1)_A$ symmetry explicitly which represents the $U(1)$ axial anomaly
\cite{'tHooft:1976up} whose strength is given
here by the constant $c$. This anomaly makes the isosinglet
Goldstone boson $\eta$ massive. 
The remaining $SU(2)$ symmetry (of three pions) stays intact
if we assume the masses of the up and down quark to be equal
so that only one diagonal generator gets a finite expectation value.
So, we choose
\begin{equation}
  \label{eq:vev}
  \langle \Phi \rangle = T_0 \phi_0 =  \frac{1}{2}\phi_0\ \openone
  \ .
\end{equation}
Finally, the last term in the Lagrangian~(\ref{eq:langrange1})
explicitly breaks chiral symmetry and makes also the pions massive. 
It resembles the mass terms
in the QCD Lagrangian where here
$H$ corresponds to the quark mass matrix and $\Phi$ to the
quark condensate. We will only deal with the case $h_0\neq 0$
and keep the $SU(2)$ isospin symmetry ($m_u=m_d$) conserved
so that $h_3=0$.

With rising temperature we expect the chiral
$SU(2)_V \times SU(2)_A \simeq SU(2)_L \times SU(2)_R$
symmetry to be restored so that the chiral partners 
($\sigma$ and $\pi$, $\eta$ and $a_0$)
become degenerate in mass. 
A violation of the axial $U(1)$ symmetry is inherent to the 
linear sigma model since its strength is directly given by
the model's parameter $c$. The restoration of this
symmetry can only be modelled in a phenomenological way
by making $c$ temperature-dependent, \emph{e.g.},
go down with rising $T$. For $c\to 0$ we expect 
the $\eta$ mass to become identical to the pion mass
above a certain temperature so that there is a full $U(2)_A$ symmetry in
the pseudoscalar sector. And, finally, all four masses are 
expected to become degenerate
at temperatures above $\langle\bar{q}q\rangle^{1/3}\approx 300$~MeV
for explicit symmetry breaking or above a critical temperature
$T_c$ in the chiral limit.

\subsection{Effective action}
\begin{figure}
  \centering
  \includegraphics[width=.8\linewidth]{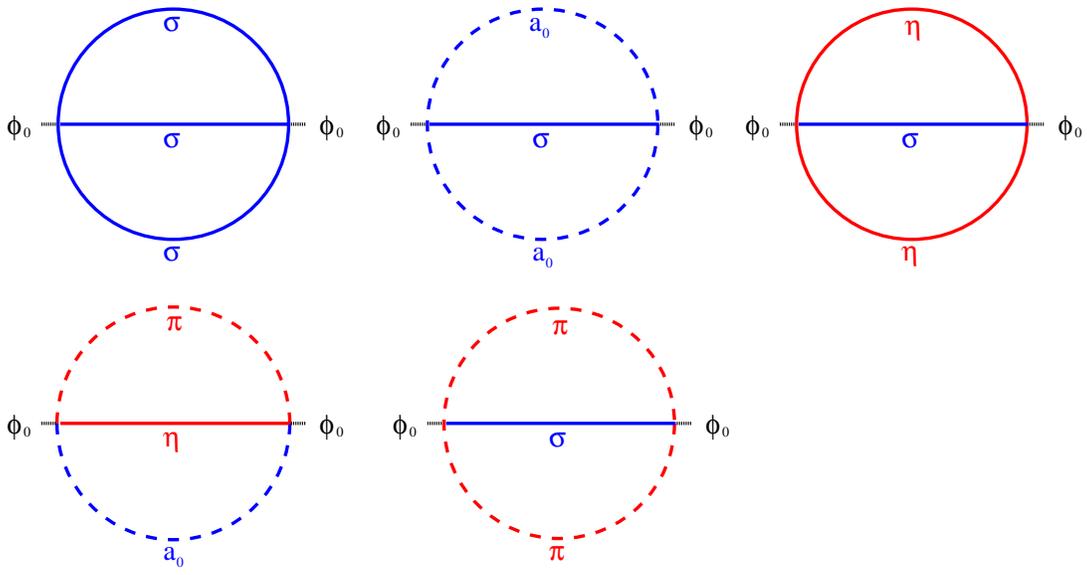}
  \caption{The sunset graphs as in Eq.~\eqref{eq:sunsets}.
  Scalars are drawn in blue, pseudoscalars in red. Dashed lines represent 
  isotriplets while solid lines are used for isosinglets.}
  \label{fig:sunsets}
\end{figure}

We compute the effective action using the 2PPI 
formalism~\cite{Smet:2001un,Verschelde:2000ta,
  Verschelde:1992bs,Baacke:2002pi}
and include all graphs up to two loops.
The reader is referred to the Appendix for details of the computation. 
Here, we will only give the final
result for the effective potential:
\begin{equation}
  V_\mathrm{eff}({M}^2,\phi_0) = V_\mathrm{cl}(\phi_0)
  + V_\mathrm{db}({M}^2,\phi_0) 
  + V_q({M}^2,\phi_0)
  \ .
\end{equation}
It is a function of the masses ${M}_\sigma^2$, ${M}_\pi^2$,
${M}_\eta^2$ and ${M}_{a_0}^2$ 
(denoted by ${M}^2$ for brevity) and the condensate $\phi_0$.
The classical part is
\begin{equation}
  \label{eq:Vcl}
  V_\mathrm{cl}(\phi_0) = \frac{1}{2}(m^2-c)\phi_0^2
  + \left( \frac{\lambda_1}{4} + \frac{\lambda_2}{8} \right)
  \phi_0^4 - h_0\,\phi_0
  \ .
\end{equation}
The quantum part $V_q$ contains all \emph{two-particle point-irreducible}
\footnote{Graphs that do not fall apart if two lines meeting at
  the same vertex are cut.}
(2PPI) graphs that can be 
made of the vertices of the shifted Lagrangian except for the
double bubbles which will be taken care of by $V_\mathrm{db}$.
The propagators within these graphs are defined by an effective
mass and have the Euclidean form
\[ G_*(p)=\frac{1}{p^2 + {M}_*^2}\ .\]
Here, we only take into account 2PPI graphs with one and two 
loops which leads to
\begin{equation}
  \label{eq:Vq}
  \begin{split}
    V_q(M^2,\phi_0) =\ 
    & \frac{1}{2} \ln\det (\partial^2+{M}_\sigma^2)
    + \frac{1}{2} \ln\det (\partial^2+{M}_\eta^2) \\
    & + \frac{3}{2} \ln\det (\partial^2+{M}_{a_0}^2)
    + \frac{3}{2} \ln\det (\partial^2+{M}_\pi^2) \\
    & + V_\mathrm{sunsets}(M^2,\phi_0)
  \end{split}
\end{equation}
The sunset contribution is given by
(note the minus sign)
\begin{equation}
  \label{eq:sunsets}
  \begin{split}
    V_\mathrm{sunsets}(M^2,\phi_0) = - \phi_0^2\, \Biggl[ & 
    \left( \lambda_1+\frac{3}{2}\lambda_2 \right)^2
    (3\, S_{\sigma\sigma\sigma}+ 3\, S_{\sigma a_0 a_0}
    + S_{\sigma\eta\eta}) \\
    & +  \left( \lambda_1+\frac{\lambda_2}{2} \right)^2
    S_{\sigma\pi\pi} 
    + \frac{3}{2}\lambda_2^2\,S_{a_0\eta\pi}
    \Biggr]\ ,
  \end{split}
\end{equation}
where $S_{ijk}$ denotes a sunset graph made of the propagators
of the particles $i$, $j$ and $k$. A graphical representation
of these contributions can be found in Fig.~\ref{fig:sunsets}.
These graphs arise from the possible
decays of $\sigma\to\pi\pi$, $\sigma\to\eta\eta$ and $a_0\to\eta\pi$.

$V_\mathrm{db}$ is the double-bubble part that receives a special
treatment in the 2PPI formalism
\begin{equation}
  \label{eq:Vdb}
  \begin{split}
    V_\mathrm{db}({M}^2,\phi_0) =\ 
    & - \left(\frac{\lambda_1}{4}+ \frac{\lambda_2}{8}\right)
    (3\, \Delta_\sigma^2 + 15\,\Delta_\pi^2 + 6\, \Delta_\sigma\Delta_\pi) \\
    & - \left(\frac{\lambda_1}{4}+ \frac{\lambda_2}{8}\right)
    (3\, \Delta_\eta^2 + 15\,\Delta_{a_0}^2 + 6\, \Delta_\eta\Delta_{a_0}) \\
    & - \left( \frac{\lambda_1}{2}+\frac{\lambda_2}{4}\right)
    \Delta_\sigma\Delta_\eta 
    - 3\,\left( \frac{\lambda_1}{2}+\frac{3}{4}\lambda_2\right)
    \Delta_{a_0}\Delta_\sigma \\
    & - 3\,\left( \frac{\lambda_1}{2}+\frac{3}{4}\lambda_2\right)
    \Delta_\pi\Delta_\eta 
    - 3\,\left( \frac{3}{2}\lambda_1+\frac{7}{4}\lambda_2\right)
    \Delta_{a_0}\Delta_\pi
    \ .
  \end{split}
\end{equation}
All quantities $\Delta$  are obtained via
\begin{equation}
  \label{eq:Delta}
  \Delta_*({M}^2,\phi_0) = 
  2\, \frac{\partial}{\partial {M}^2_*}V_q({M}^2,\phi_0)
\end{equation}
where $*$ stands for $\sigma$, $a_0$, $\eta$ or $\pi$
and $M^2$ is a short-hand notation for all four masses.
So, these quantities are explicit function of the
mass matrix ${M}^2$ and the condensate $\phi_0$.

\subsection{Equations of motion}
The mass gap equations in the 2PPI effective action formalism
are given by
\[ \frac{\partial V_\mathrm{eff}}{\partial M^2_*} = 0
\qquad\text{with}\quad *=\sigma,a_0,\eta,\pi\]
and read explicitly
 \begin{subequations}
   \label{eq:gap_u2}
   \begin{eqnarray}
     \begin{split}
       {M}^2_\sigma = & m^2-c 
       + 3 \left( \lambda_1 + \frac{\lambda_2}{2}\right) \phi_0^2 
       + 3\,\left( \lambda_1 +  \frac{\lambda}{2}\right)
       (\Delta_\sigma+\Delta_{\pi}) \\
       & + 3\,\left( \lambda_1 + \frac{3}{2}\lambda_2\right)
       \Delta_{a_0} 
       + \left( \lambda_1 + \frac{\lambda_2}{2}\right)\,\Delta_\eta
     \end{split}\\
     \begin{split}
       {M}^2_\pi = & m^2-c 
       + \left( \lambda_1 + \frac{\lambda_2}{2}\right) \phi_0^2 
       + \left( \lambda_1 +  \frac{\lambda}{2}\right)
       (\Delta_\sigma+ 5\,\Delta_{\pi}) \\
       & + \left( \lambda_1 + \frac{3}{2}\lambda_2\right)
       \Delta_{\eta} 
       + \left( 3\,\lambda_1 + \frac{7}{2}\lambda_2\right)\,\Delta_{a_0}
     \end{split}\\
     \begin{split}
       {M}^2_\eta =& m^2+c 
       + \left( \lambda_1 + \frac{\lambda_2}{2}\right) \phi_0^2 
       + \left( \lambda_1 +  \frac{\lambda}{2}\right)
       3\,(\Delta_\eta+\Delta_{a_0}) \\
       & + \left( \lambda_1 + \frac{\lambda_2}{2}\right)
       \Delta_{\sigma} 
       + 3\,\left( \lambda_1 + \frac{3}{2}\lambda_2\right)\,\Delta_{\pi}
     \end{split} \\
     \begin{split}
       {M}^2_{a_0} =& m^2+c
       + \left( \lambda_1 + \frac{3}{2}\lambda_2\right) \phi_0^2 
       + \left( \lambda_1 +  \frac{\lambda}{2}\right)
       (\Delta_\eta + 5\, \Delta_{a_0}) \\
       & + \left( \lambda_1 + \frac{3}{2}\lambda_2\right)
       \Delta_{\sigma} 
       + \left( 3\,\lambda_1 + \frac{7}{2}\lambda_2\right)\,\Delta_{\pi}
       \ .
     \end{split}
   \end{eqnarray}
 \end{subequations}
The formalism is made such that these equations
 resemble those of the \HF\ approximation
with the decisive difference that here $\Delta_*$ is not a single bubble
but calculated from Eq.~(\ref{eq:Delta}). Neglecting the sunsets 
contributions to $V_q$ in Eq.~(\ref{eq:Delta}) would 
reduce all quantum corrections to simple
bubbles; in this way the \HF\ approximation is regained.
Solving Eqs.~\eqref{eq:gap_u2} for all four quantum corrections
one finds a (quite lengthy) expression for each quantum correction in 
terms of all four 
masses and the condensate. 
The equation of motion for the condensate 
\[ \frac{\partial}{\partial \phi_0} V_\mathrm{eff}({M}^2,\phi_0)
= 0\]
can be put in
a very easy form by simplifying $V_\mathrm{db}({M}^2,\phi_0)
+V_\mathrm{cl}(\phi_0)$:
\begin{equation}
  \label{eq:condensate}
  h_0 = 
  {M}^2_\sigma\, \phi_0- (2\lambda_1 + \lambda_2)\, \phi_0^3
  + \frac{\partial}{\partial \phi_0} V_\mathrm{sunsets}
  \ .
\end{equation}
Neglecting the sunsets this equation is equivalent to the
one found in the \HF\ approximation~\cite{Roder:2003uz}.

\section{The $O(N)$ linear sigma model}
\label{sec:on-linear-sigma}
\subsection{Langrangian}
The linear sigma model with an $O(4)$ symmetry is obtained from
the $U(2)_V\times U(2)_A$ model in the limit of infinite anomaly
$c\to\infty$ with fixed $(m^2-c)\to m^2_{O(4)}$. 
The masses of both the $\eta$ and the $a_0$ mesons become infinite
and thus these two mesons drop out of the spectrum.
The coupling is renamed to 
$\lambda\equiv\left(\lambda_1 + \frac{\lambda_2}{2}\right)$, and
$\sigma$ and the three pions now share one $O(4)$ multiplet.
Extending the isospin symmetry from four to $N$ dimensions we can
now write down the
well-known Lagrangian of the $O(N)$ linear sigma model
\begin{equation}
  \label{eq:Lo4}
  \mathscr{L}[\Phi] = \frac{1}{2} \left( \partial_\mu \Phi_i \right)^2
  - \frac{1}{2} m^2 \Phi_i^2 - \frac{\lambda}{4} \left(\Phi_i^2\right)^2
  + h_i\, \Phi_i
  \quad\text{with}\quad i=1,\dots, N
\end{equation}
which has been studied 
extensively~\cite{Lenaghan:1999si,Baacke:2002pi,Baacke:2003dk,
  Petropoulos:1998gt,Dolan:1973qd,Chiku:1998kd,Nemoto:1999qf,
  Smet:2001un,Verschelde:2000ta,Andersen:2004ae}.
In this article we will extend the analysis performed
earlier~\cite{Baacke:2002pi,Baacke:2003dk} to the case of 
explicit symmetry breaking and realistic values of the 
parameters.

\subsection{Equations of motion}
The equations of motion are obtained from the 2PPI effective 
potential where the vacuum expectation value 
is set to be
\[ \langle \Phi_i \rangle = \phi_0\,\delta_{0i}\ ,\]
so that the 2PPI effective action 
reads~\cite{Baacke:2002pi,Verschelde:2000ta}
\begin{equation}
  \label{eq:effpot_o4}
  \begin{split}
    V_\mathrm{eff}(\phi_0; M_\sigma^2, M_\pi^2) =\ & 
    \frac{1}{2}M_\sigma^2\phi_0^2 - \frac{\lambda}{2}\phi_0^4
    - h_0\,\phi_0
    + \frac{m^2}{2\lambda\,(N+2)} \left[ M_\sigma^2 + (N-1) 
      M_\pi^2 \right] \\
    & - \frac{1}{8\lambda\,(N+2)}\biggl[ (N+1)\,M_\sigma^4 
      + 3\,(N-1) M_\pi^4 \\
    &  - 2\,(N-1)\, M_\sigma^2 M_\pi^2 + 2N\lambda^2 m^4
    \biggr]
    + V_q(\phi_0; M_\sigma^2, M_\pi^2)
    \ .
  \end{split}
\end{equation}
The quantum corrections consist of the following
one- and two-loop terms
\begin{equation}
  \label{eq:Vq-o4}
  \begin{split}
    V_q(\phi_0; M_\sigma^2, M_\pi^2) =\ & 
    \frac{1}{2} \ln\det(\partial^2+{M}_\sigma^2) 
    + \frac{N-1}{2}\ln\det(\partial^2+{M}_\pi^2)\\
    & - \lambda^2\phi_0^2\left[
      3\,S_{\sigma\sigma\sigma} + (N-1)\, S_{\sigma\pi\pi}
    \right]\ .
  \end{split}
\end{equation}
The mass gap equations follow from the stationarity conditions
\[ \frac{\partial}{\partial M_\sigma} V_\mathrm{eff} = 0
\quad\text{and}\quad
\frac{\partial}{\partial M_\pi} V_\mathrm{eff} = 0 \ .
\]
They read
\begin{subequations}
  \label{eq:gap_o4}
  \begin{eqnarray}
    {M}_\sigma^2 &=& m^2 + 3\,\lambda\phi_0^2 +
    \lambda\left[ 3\,\Delta_\sigma + (N-1)\, \Delta_\pi \right] \\
    {M}_\pi^2 &=& m^2 + \lambda\phi_0^2 +
    \lambda\left[ \Delta_\sigma + (N+1)\, \Delta_\pi \right]\ ,
  \end{eqnarray}
\end{subequations}
where all quantum corrections $\Delta$ are explicit 
functions of both the condensate $\phi_0$ and the
masses defined as 
\[ \Delta_i({M}_\sigma, {M}_\pi; \phi_0) = 
2\, \frac{\partial V_q}{\partial {M}_i^2}\ .
\]
The equation for the condensate has the same structure as
the one in the \Utwo\ model, Eq.~\eqref{eq:condensate},
\begin{equation}
  \label{eq:cond_o4}
  h_0 = \left\{ {M}_\sigma^2 - 2\lambda\,\phi_0^2
    - 2\,\lambda^2 \left[
      3\,S_{\sigma\sigma\sigma} + (N-1)\, S_{\sigma\pi\pi}
      \right] 
    \right\}\, \phi_0\ .
\end{equation}
In the following we will only investigate the case $N=4$.


\section{Numerical Results}
\label{sec:numerical-results}
\subsection{Parameter fixing and loop graphs at finite 
  temperature}

The parameters in both models are fixed such that at $T=0$
the values of all masses are equal to the
values in the Particle Physics Booklet~\cite{Eidelman:2004wy}, 
cf.~Table~\ref{tab:masses}, where we choose the mass of the
$\sigma$ meson to be 600~MeV. 
The value of the condensate $\phi_0$ is related to the
mesons decay constants $f_a$ and determined by the 
PCAC (partial conservation of axial vector current) hypothesis
\[ f_a = d_{aa0}\ \phi_0 \equiv \phi_0\ . \] This fixes the condensate to 
$\phi_0 = f_\pi$ because $d_{aa0}=1$, so 
all decay constants are the same.

For the two models of this paper
the fixing can be done in a unique way because there are
as many equations of motion as parameters. At tree-level it could be done
even in the chiral limit (with $h_0$ fixed to zero) since
the equation for the condensate coincides with the one for
the pion mass. 
Problems occur if one wants to include terms that contain a
renormalization scale because, first, in the chiral limit
there is only one possible value for this scale where the parameters 
can be fixed and, second, for explicit symmetry breaking
the temperature-dependence of the
condensate and the masses is varying with the renormalization
scale~\cite{Lenaghan:1999si}. 
So, the system only gets an extra parameter and all
quantities are logarithmically dependent on this scale 
which makes the results somewhat arbitrary. 
Furthermore, terms originating from
renormalization can be such that the gap equations are not
solvable above a certain 
temperature~\cite{Baym:1977qb,Chiku:1998kd,Bardeen:1986td}.
Lenaghan and Rischke~\cite{Lenaghan:1999si} have also shown
that, in the $O(N)$ model, there is no qualitative difference
whether one includes the finite renormalization terms or not.
In order to get rid of this extra parameter
we take the phenomenological approach proposed 
before~\cite{Roder:2003uz,Lenaghan:1999si,Lenaghan:2000ey} and set all 
finite terms arising from regularization equal to zero which makes all
quantum corrections only play a role at finite temperature.
So, effectively the parameters are fixed at tree-level.
The resulting values
can be found in Tables~\ref{tab:parametersU2} and~\ref{tab:parametersO4}.

Neglecting finite terms from renormalization
the one-loop graphs at finite temperature
--- the boson determinants in the effective action and
the single bubble (or tadpole) $\mathcal{B}$ --- 
 are given by the following equations~\cite{Baacke:2002pi}
\begin{subequations}
  \begin{eqnarray}
    \label{eq:lndet}
    \ln \det (\partial^2+M_i^2)
    &=& \frac{T}{\pi^2} \int_0^\infty\!\!\! dp\ 
    {\bf p}^2\, \ln\left[ 1-e^{-E({\bf p})/T} \right] \\
    \label{eq:bubble}
    \mathcal{B}_i &=& 
    \frac{\partial}{\partial M_i^2}\ln \det (\partial^2+M_i^2)
    = \frac{1}{2\,\pi^2} \int_0^\infty\!\!\! dp\
    \frac{{\bf p}^2}{E_i(\bf p)}\, n_i({\bf p})\ ,
  \end{eqnarray}
\end{subequations}
where $E_i({\bf p})=\sqrt{{\bf p}^2+M_i^2}$ and $n_i$
is the Bose-Einstein distribution
\begin{equation}
  \label{eq:bose-einstein}
  n_i({\bf p})=\frac{1}{e^{E_i({\bf p})/T}-1}
  \ .
\end{equation} 
The sunset graph with three different masses $M_i$, $M_j$ and $M_k$
is composed of three parts in each of which two of the three particles
are taken from the heat bath
\begin{equation}
  \label{eq:sunset_therm}
  S_{ijk} = \mathcal{S}^{(2)}_{(ij)k} + 
  \mathcal{S}^{(2)}_{(ki)j} + 
  \mathcal{S}^{(2)}_{(jk)i}\ .
\end{equation}
Here, $\mathcal{S}^{(2)}_{(ij)k}$ is a sunset graph with
$i$ and $j$ being thermal lines \cite{Baacke:2002pi}
\begin{equation}
  \label{eq:S2}
  \mathcal{S}^{(2)}_{(ij)k} = \frac{1}{32\,\pi^4}
  \iint_0^\infty \!\!\! dp_i\,dp_j\ 
  \frac{p_i p_j}{E_i E_j}\, n_i({\bf p}_i)\,n_i({\bf p}_i)
  \,\ln\left|\frac{Y_{(ij)k}^+}{Y_{(ij)k}^-}\right|
\end{equation}
with 
\[ Y_{(ij)k}^\pm = \left[ (E_i+E_j)^2-(E_{(ij)k}^\pm)^2 \right]\cdot
   \left[ (E_i-E_j)^2-(E_{(ij)k}^\pm)^2 \right] 
\] and \[ E_{(ij)k}^\pm = \sqrt{(p_i\pm p_j)^2 + M_k^2}\ .\]

\begin{table}
  \centering
  \begin{tabular}{ccccc}
    \hline\hline
    & \multicolumn{4}{c}{mass in MeV} \\
    & \multicolumn{2}{c}{explicit symmetry breaking} &
    \multicolumn{2}{c}{chiral limit} \\
    particle & with anomaly & without anomaly 
    & with anomaly & without anomaly\\
    \hline
    $\sigma$ & 600.0 & 600.0 & 600.0 & 600.0\\
    $\pi$ & 139.6 & 139.6 & 0 & 0\\
    ${a_0}$ & 984.7 & 984.7 & 984.7 & 984.7 \\
    $\eta$ & 547.3 & 139.6 & 547.3 & 0 \\
    \hline\hline
  \end{tabular}
  \caption{Meson masses in the linear sigma models with and without
    axial anomaly for explicit symmetry breaking and in the chiral limit. 
    Note that, of course, there is no $a_0$ and $\eta$
    meson in the $O(4)$ model.}
  \label{tab:masses}
\end{table}

\begin{table}
  \centering
  \begin{tabular}{ccccc}
    \hline\hline
    & \multicolumn{2}{c}{explicit symmetry breaking}
    & \multicolumn{2}{c}{chiral limit} \\
    parameter 
    & with anomaly & without anomaly &
    with anomaly & without anomaly\\
    \hline
    $m^2$ & $-(103.78$~MeV)$^2$ & $-(388.34$~MeV)$^2$ &
    $-(173.87)$~MeV)$^2$ & $-(424.26$~MeV)$^2$ \\
    $c$ & (374.22~MeV)$^2$ &  0  &
    (387.00~MeV)$^2$ &  0 \\
    $h_0$ & (121.60~MeV)$^3$ &  (121.60~MeV)$^3$
    & 0 & 0\\
    $\lambda_1$ & $-19.30$ &  $-35.70$ & 
     $-19.14$ &  $-37.63$\\
    $\lambda_2$ & 78.49 & 111.29 & 82.73 & 119.71\\
    $\phi_0$ & 92.4~MeV & 92.4~MeV & 90~MeV & 90~MeV\\
    \hline\hline
  \end{tabular}
  \caption{Parameters in the \Utwo\ model for masses as in 
    Table~\ref{tab:masses}.}
  \label{tab:parametersU2}
\end{table}

\begin{table}
  \centering
  \begin{tabular}{ccc}
    \hline\hline
    parameter & explicit symmetry breaking & chiral limit\\
    \hline
    $m^2$ & $-(388.34$~MeV)$^2$ & $-(424.26$~MeV)$^2$\\
    $h_0$ & (121.6~MeV)$^3$ & 0 \\
    $\lambda$ & $19.94$ & $22.22$\\
    $\phi_0$ & 92.4~MeV & 90~MeV\\
    \hline\hline
  \end{tabular}
  \caption{Parameters in the $O(4)$ model for masses as in 
    Table~\ref{tab:masses}.}
  \label{tab:parametersO4}
\end{table}

\subsection{$O(4)$ model}
\label{sec:numeric_on}
\begin{figure}
  \centering
  \includegraphics[width=.8\linewidth]{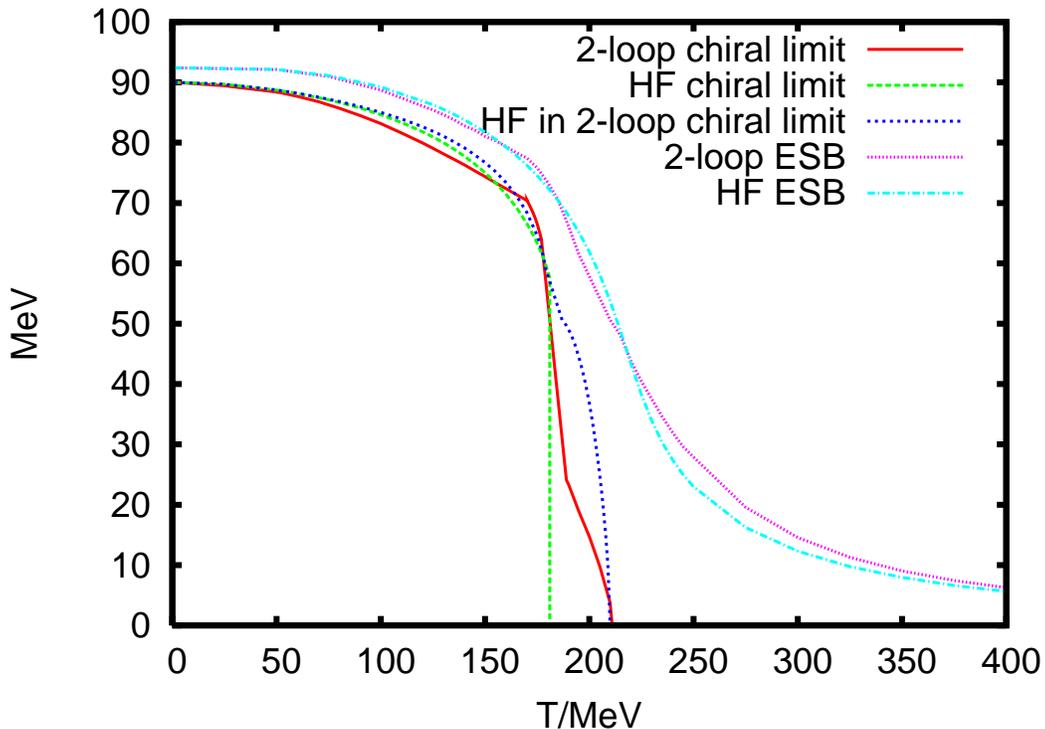}
  \caption{Temperature dependence of the condensate
    in the $O(4)$ model. Comparison of the 
    two-loop approximation with the \HF\ approximation (HF).}
  \label{fig:phi_o4_esb_600}
\end{figure}

For a given temperature $T$ 
we let $N=4$, fix the value of $\phi_0$ and then numerically extremize
the effective potential in Eq.~(\ref{eq:effpot_o4}) with respect to
$M_\sigma$ and $M_\pi$. Thereby, we
obtain a 1PI potential that is only a function of $\phi_0$.
Using the equation of motion for the condensate~(\ref{eq:cond_o4})
we eventually find the temperature-dependent value of the
order parameter $f_\pi(T)$. This quantity is plotted in
Fig.~\ref{fig:phi_o4_esb_600}.
For comparison we also plot the result for the \HF\
approximation and for a simplified two-loop approximation
(HF in 2-loop) which consists in solving the gap equations~\eqref{eq:gap_o4}
in the \HF\ approximation but
the condensate equation~\eqref{eq:cond_o4} with the sunset
contributions. In the chiral limit the transition temperature
is almost the same as in the true two-loop approximation though
the condensate drops faster at lower temperatures and then
eventually approaches zero at $T\approx 210$~MeV.
In the chiral limit of the \HF\ approximation,
 there is a first-order phase transition at
$T_c\approx 181$~MeV whereas the two-loop approximations 
(both the true and the simplified one)
correctly reproduce a second-order transition though
at a higher temperature of about 210~MeV.

For explicit symmetry breaking the simplified two-loop
approximation results lie on top of the \HF\
results, only the true two-loop approximation
yields slight deviations. There, the
condensate drops faster at temperatures below 200~MeV and decreases 
more slowly for temperatures above 250~MeV 
than but in all approximations the crossover temperature ---
the one where the slope of $f_\pi(T)$ is largest ---
is about 225~MeV.

\begin{figure}
  \centering
  \subfigure[Explicit symmetry breaking.]{
    \includegraphics[width=.8\linewidth]{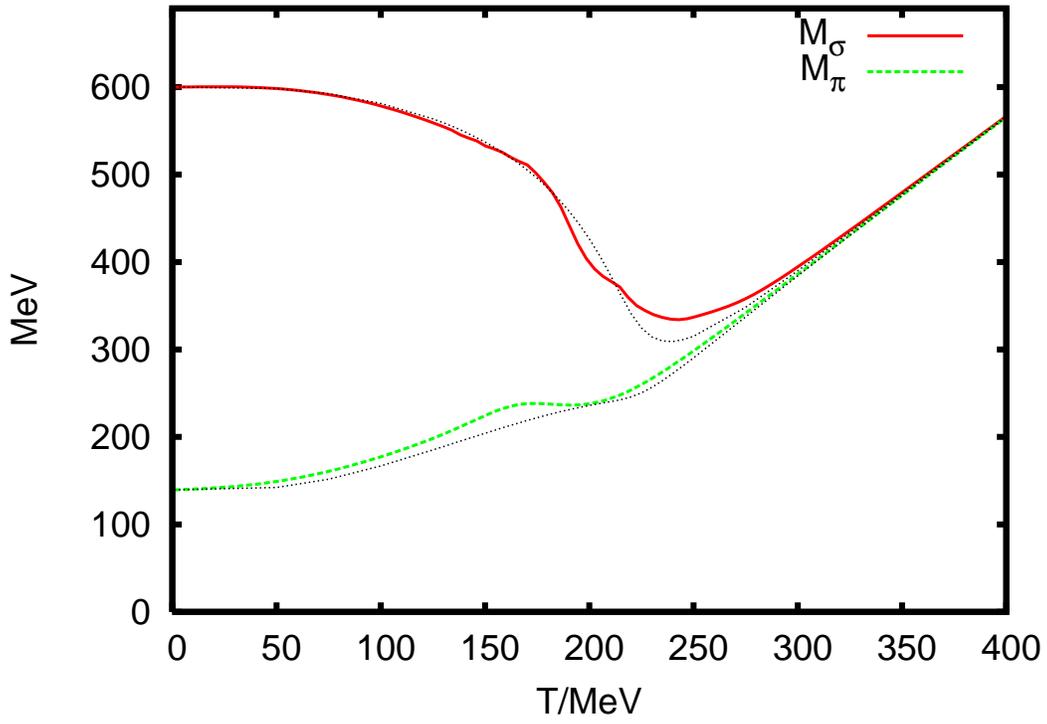}
    \label{fig:o4_masses_esb}
  }
  \subfigure[Chiral limit.]{
    \includegraphics[width=.8\linewidth]{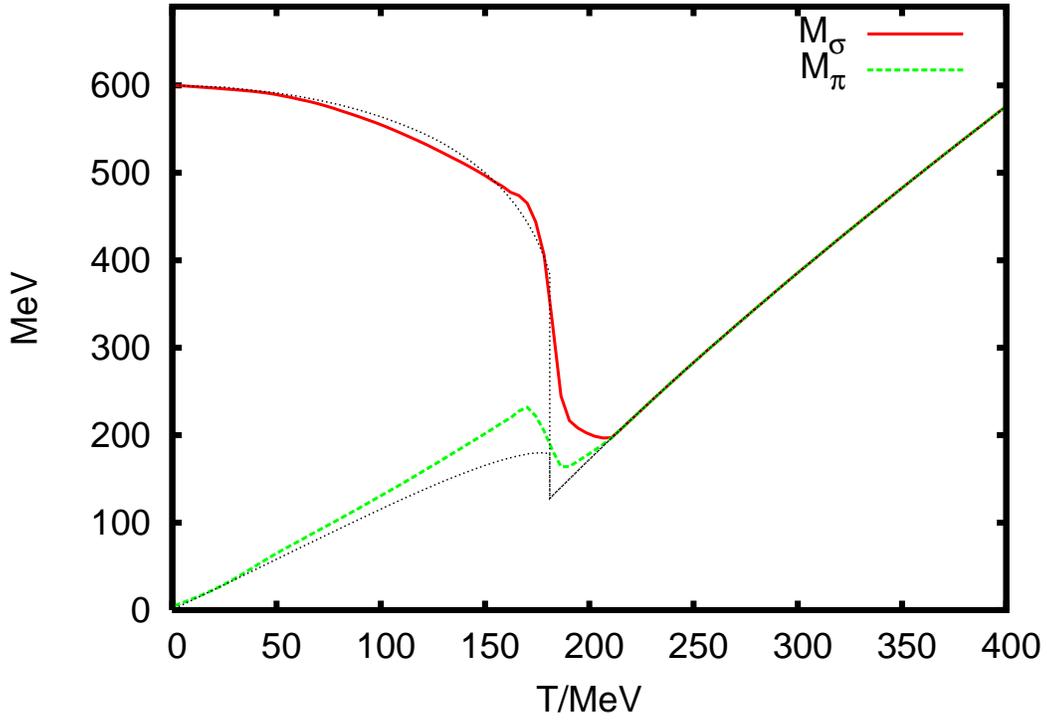}
    \label{fig:o4_masses_ssb}
  }
  \caption{Temperature dependence of the masses
    in the $O(4)$ model. Comparison of the 
    two-loop approximation with \HF\ (dotted).}
  \label{fig:o4_masses}
\end{figure}

The temperature dependence of the $\sigma$ and pion mass is
displayed in Fig.~\ref{fig:o4_masses}. In the chiral
limit [Fig.~\ref{fig:o4_masses_esb}] the $\sigma$ mass
behaves similarly in both approximations while the pion mass
first increases before it slightly drops at $T\approx 195$~MeV 
(180~MeV in the chiral limit). Beyond
temperatures of about 200~MeV the pion mass in both approximations is
almost the same. In the vicinity of 300~MeV both masses become 
identical, a sign for the restoration of chiral symmetry. There, the
temperature (to the third power) is equal to the value of the
chiral quark condensate (see above).
In the chiral limit [Fig.~\ref{fig:o4_masses_ssb}] the pion
mass first grows stronger in the two-loop approximation than in
\HF, then it slightly drops before the critical temperature is
reached and the masses become the same.

Moreover, there is a violation of
Goldstone's theorem at finite temperature in the chiral limit
because even for temperatures lower than the critical one,
$M_\pi$ is not equal to zero although the symmetry
is spontaneously broken. This phenomenon can also be
found in earlier results
on the $O(N)$ model~\cite{Baacke:2002pi,Verschelde:2000ta,Nemoto:1999qf,
  Roder:2003uz}. 
We will comment on this in Sec.~\ref{sec:conclusions-outlook}.

\subsection{\Utwo\ model}
The procedure performed for the $O(N)$ model, 
cf.~Sec.~\ref{sec:numeric_on}, cannot be done for the 
\Utwo\ model because it turns out that the
potential $V=V_\mathrm{cl}+V_\mathrm{db}$ 
has only a saddle point with respect
to the four masses instead of a local extremum.

\begin{figure}
  \centering
  \includegraphics[width=.8\linewidth]{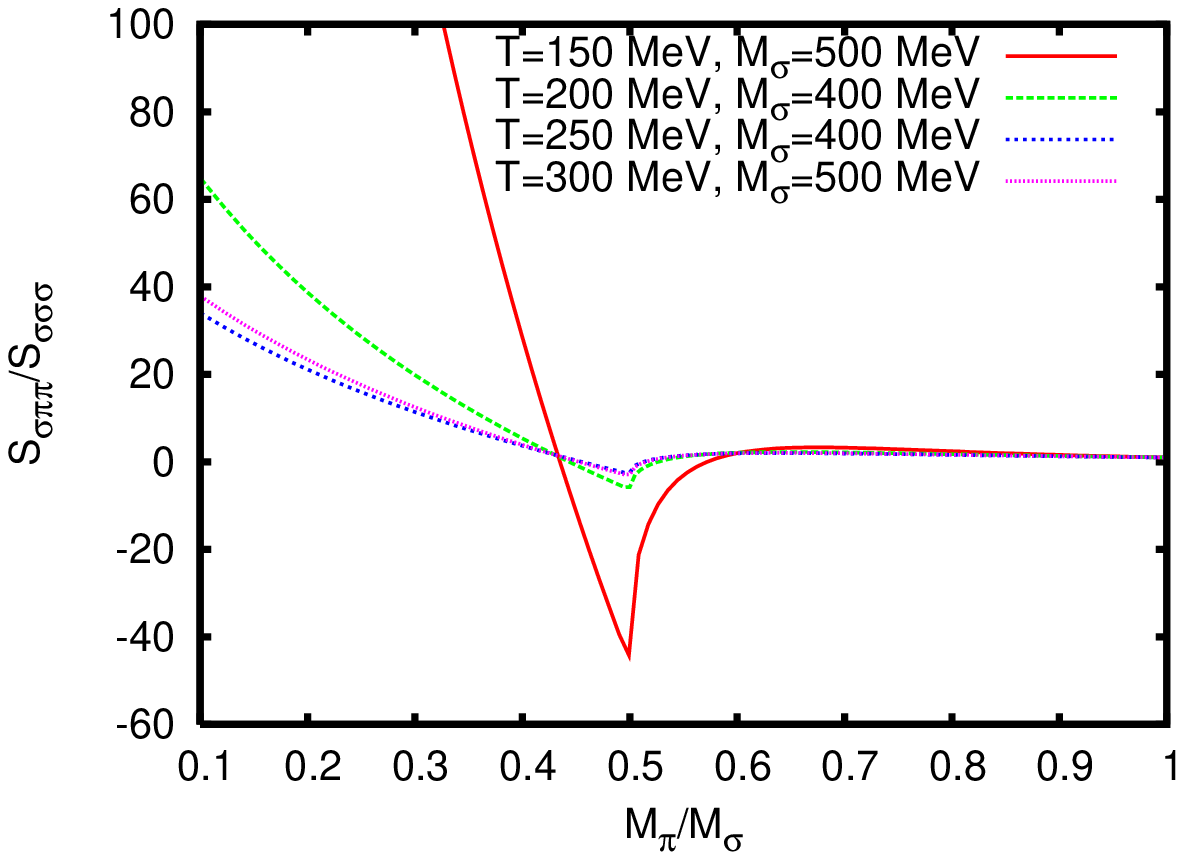}
  \caption{The ratio $S_{\sigma\pi\pi}/S_{\sigma\sigma\sigma}$ 
    for $M_\sigma=500$~MeV at $T=150$~MeV and $T=300$~MeV,
    and $M_\sigma=400$~MeV at $T=200$~MeV and $T=250$~MeV.}
  \label{fig:sunset_150}
\end{figure}

Solving the gap equations~\eqref{eq:gap_u2} is a cumbersome procedure because
they contain derivatives of the sunset graph with respect to
a mass. And 
in the vicinity of the decay threshold of the particles involved in a
sunset graph, \textit{e.g.} $M_\sigma\simeq 2\,M_\pi$,
the sign of the derivative of the sunset graph 
quickly changes (see Fig.~\ref{fig:sunset_150})
which results in a numerically unstable
behavior in this region. To avoid this trouble a simplified
approximation is used (cf. Sec.~\ref{sec:numeric_on}): we solve the mass 
gap equations~(\ref{eq:gap_u2}) in
the \HF\ approximation, \textit{i.e.}, the quantum corrections
$\Delta$ to the masses only consist of single bubbles given
by Eq.~\eqref{eq:bubble}.
We substitute these masses into the two-loop equation 
for the condensate~\eqref{eq:condensate}
to obtain the temperature-dependent order parameter or decay constant
 $f_\pi(T)$ and, for simplicity,  call
this approximation ``two-loop'' from now 
on.

\subsubsection{Model with axial $U(1)$ anomaly}
\label{sec:u2_results_ano}

\begin{figure}
  \centering
  \includegraphics[width=.8\linewidth]{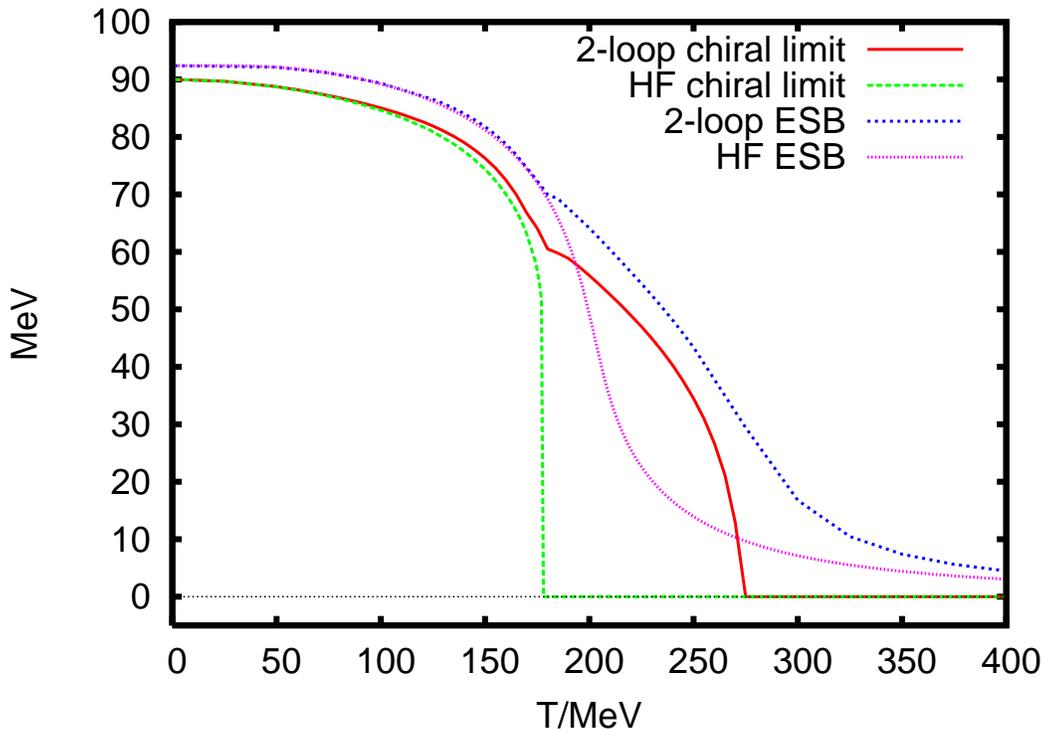}
  \caption{Temperature dependence of the condensate
    in the \Utwo\ model with an axial $U(1)$ anomaly. Comparison of the 
    two-loop approximation with \HF\ (HF) 
    for explicit symmetry breaking (ESB)
    and in the chiral limit.}
  \label{fig:u2_ano_condensate}
\end{figure}

\begin{figure}
  \centering
  \subfigure[Explicit symmetry breaking.]{
    \includegraphics[width=.8\linewidth]{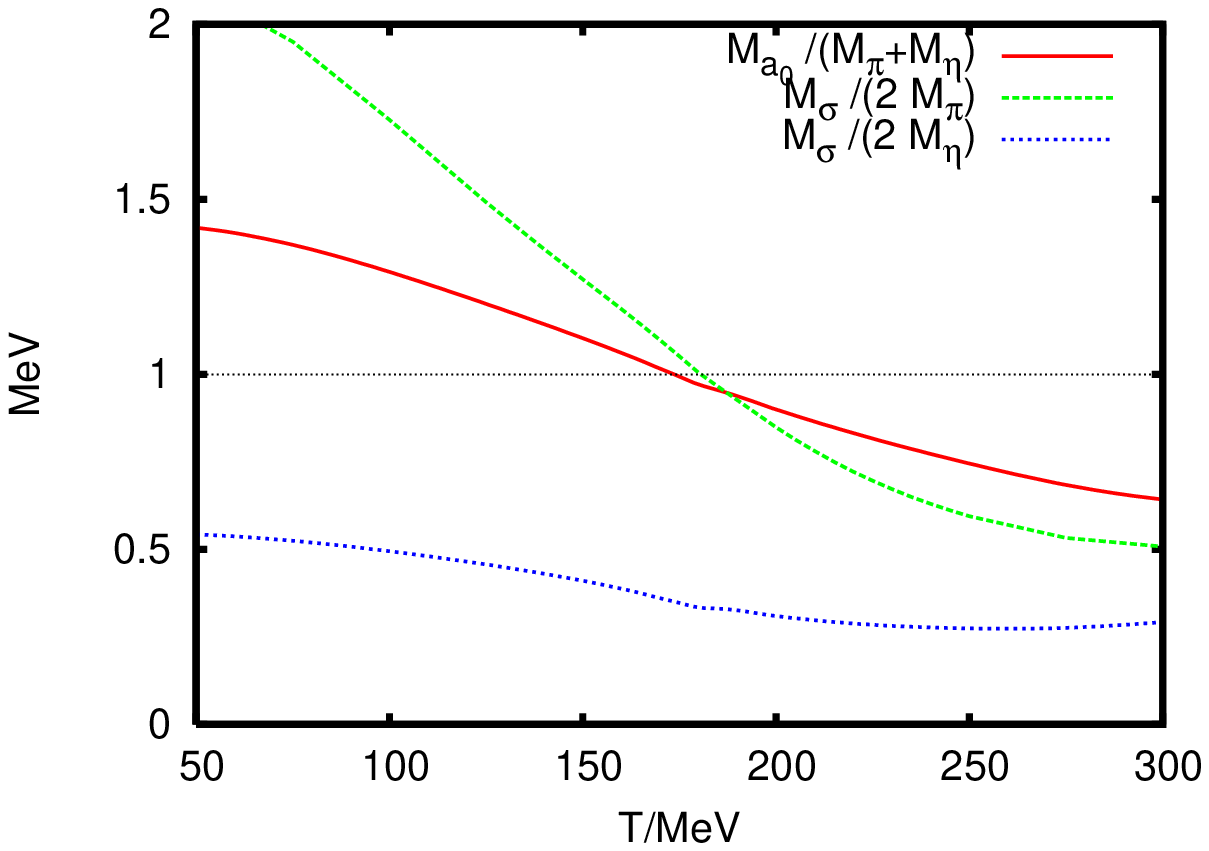}}

  \subfigure[Chiral limit.]{
    \includegraphics[width=.8\linewidth]{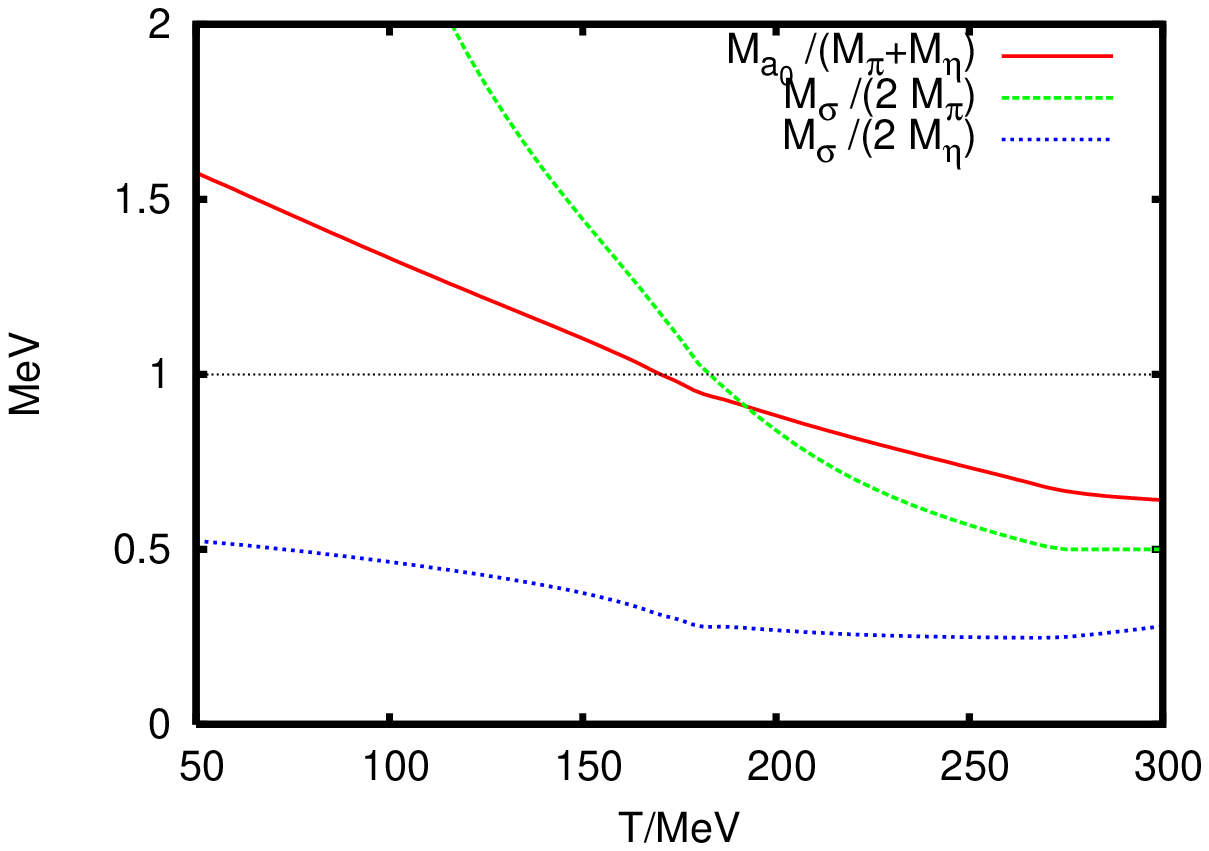}}
  \caption{Thresholds for the decays $\sigma\to \pi\pi$,
    $a_0\to \eta\pi$ and $\sigma\to\eta\eta$
    in the \Utwo\ model with axial anomaly as.}
  \label{fig:u2_ano_thresh}
\end{figure}

In Fig.~\ref{fig:u2_ano_condensate} we show the temperature
dependence of the condensate in the \Utwo\ model with
an axial anomaly.
In the chiral limit of the \HF\ approximation 
there is a first-order phase transition
at $T_c\approx 178$~MeV where the condensate discontinuously drops
down from about 52~MeV to zero. In the two-loop approximation
we find a second-order transition at $T_c\approx 272$~MeV.
So, the two-loop approximation correctly reproduces the
order of the phase transition obtained 
from universality class arguments~\cite{Pisarski:1983ms},
though the critical
temperature is about 100~MeV higher than suggested by QCD lattice
calculations~\cite{Karsch:2001cy,Laermann:2003cv}. 
The artificial first-order transition
in the \HF\ approximation has been discovered before by R\"oder 
\etal\ for a $\sigma$ mass of 400~MeV~\cite{Roder:2003uz}. 
For both explicit symmetry breaking and in the chiral limit
the order parameter decreases more slowly
with rising temperature 
in the two-loop approximation than in the \HF\ approximation
and exhibits some ``bumps'' in the curve.
The reason for that behavior is a decay threshold effect caused
by the sunset graphs 
[cf. Eq.~\eqref{eq:sunsets} and Fig.~\ref{fig:sunset_150}].
To check where the thresholds are crossed we plot the mass ratios 
for the decays $\sigma\to\pi\pi$, 
$a_0\to\eta\pi$ and $\sigma\to\eta\eta$
in Fig.~\ref{fig:u2_ano_thresh}.

The curve of the order parameter 
in the two-loop approximation deviates from the one of the 
\HF\ approximation at a 
temperature of about 175-180~MeV. This is the region where the
thresholds of the decays $\sigma\to\pi\pi$ and $a_0\to\eta\pi$
 are crossed (see Fig.~\ref{fig:u2_ano_thresh}). The sunsets
become larger with rising temperature but also decrease with
masses approaching the threshold from below. The latter behavior
suddenly changes at the threshold and the sunsets suddenly grow
which causes the immediate deviation from the \HF\ results. Looking 
at the equation for the condensate~\eqref{eq:condensate} one
can roughly conclude that a larger value of the condensate is
needed to compensate for the contribution from the sunset terms
\footnote{This argument disregards the fact that all quantities in
  that equation depend on the condensate because we deal with
  a self-consistent resummation scheme.}.

Comparing the prefactors of the different sunsets in 
Eq.~\eqref{eq:sunsets} using the numerical values of the couplings from
Table~\ref{tab:parametersU2} we notice that, in the case of a finite 
$U(1)_A$ anomaly,  the 
contributions from $S_{\sigma\eta\eta}$ and $S_{a_0 \eta\pi}$ are almost of the
same size whereas $S_{\sigma\pi\pi}$ is smaller by a factor of eight.
Furthermore, the sunset term has dimension two and thus scales
with temperature and the masses of the particles within.
So, considering the prefactors and the different values for the masses ---
where $M_{a_0}$ and $M_\eta$ are the two largest ones --- 
the deviation from the \HF\ 
approximation is dominated by a threshold effect
of the decay $a_0\to\eta\pi$.

\begin{figure}
  \centering
  \subfigure[Explicit symmetry breaking]{
    \includegraphics[width=.8\linewidth]{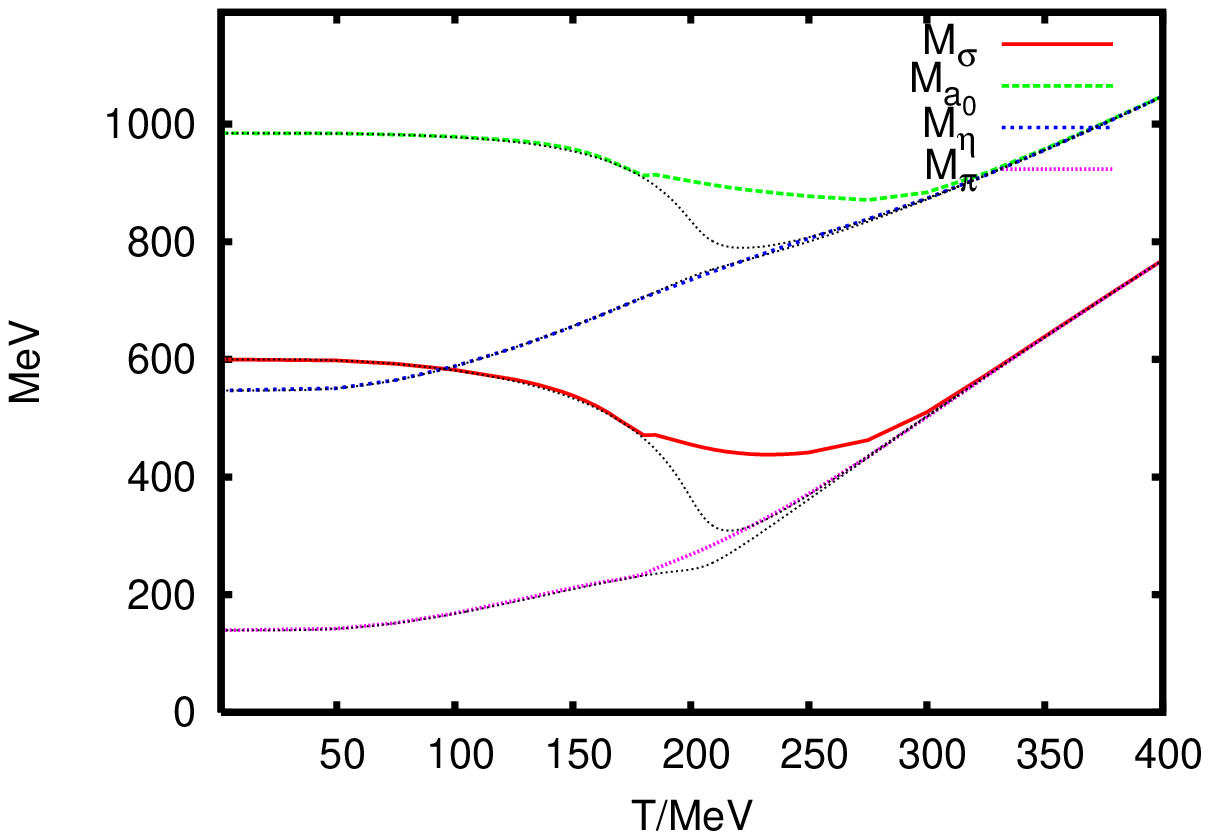}
    \label{fig:u2_esbano_masses}
  }

  \subfigure[Chiral limit.]{
    \includegraphics[width=.8\linewidth]{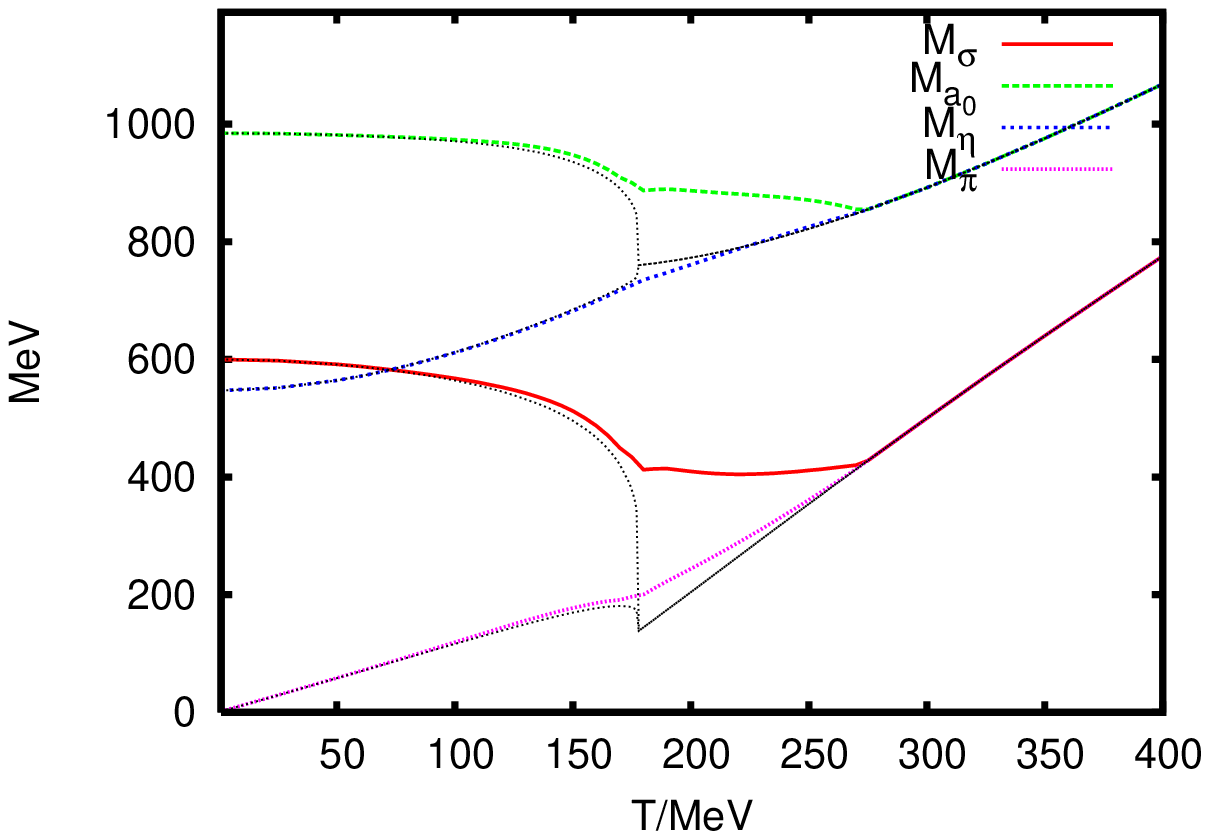}
    \label{fig:u2_ssbano_masses}
  }
  \caption{Masses in the \Utwo\ model with axial anomaly 
    as functions of temperature. 
    Comparison of the two-loop approximation
     with \HF\ (dotted lines).}
  \label{fig:u2_ano_masses}
\end{figure}

The temperature-dependent masses for the \Utwo\ model 
are displayed in Fig.~\ref{fig:u2_ano_masses}.
Especially the masses of the 
scalar mesons $\sigma$ and $a_0$ behave differently in the 
two-loop approximation
than in \HF\ whereas the masses of both pseudoscalar mesons exhibit no
qualitative difference in their temperature dependence between 
the two approximations. The observed
deviations are due to the aforementioned threshold effects through
sunset graphs. At a temperature of about 300~MeV 
(or beyond $T_c \approx 272$~MeV in the
chiral limit) the masses of
the chiral partners become identical so that chiral symmetry
is restored. The $U(1)_A$ symmetry remains broken since the 
parameter $c$ is not a function of temperature. The gap between
the mass squares of the isospin partners $\eta$ and $\pi$ 
(or $\sigma$ and $a_0$) remains 
equal to $2c$ according to Eq.~\eqref{eq:gap_u2}.
In the \Utwo\ model as well, the pion mass is finite in the chiral limit
even for $T<T_c$ which was discovered in the \HF\ approximation
earlier~\cite{Roder:2003uz}. We will give a statement concerning
a possible violation of Goldstone's theorem in 
Sec.~\ref{sec:conclusions-outlook}.

\subsubsection{Model without axial $U(1)$ anomaly}
\begin{figure}
  \centering
  \includegraphics[width=.8\linewidth]{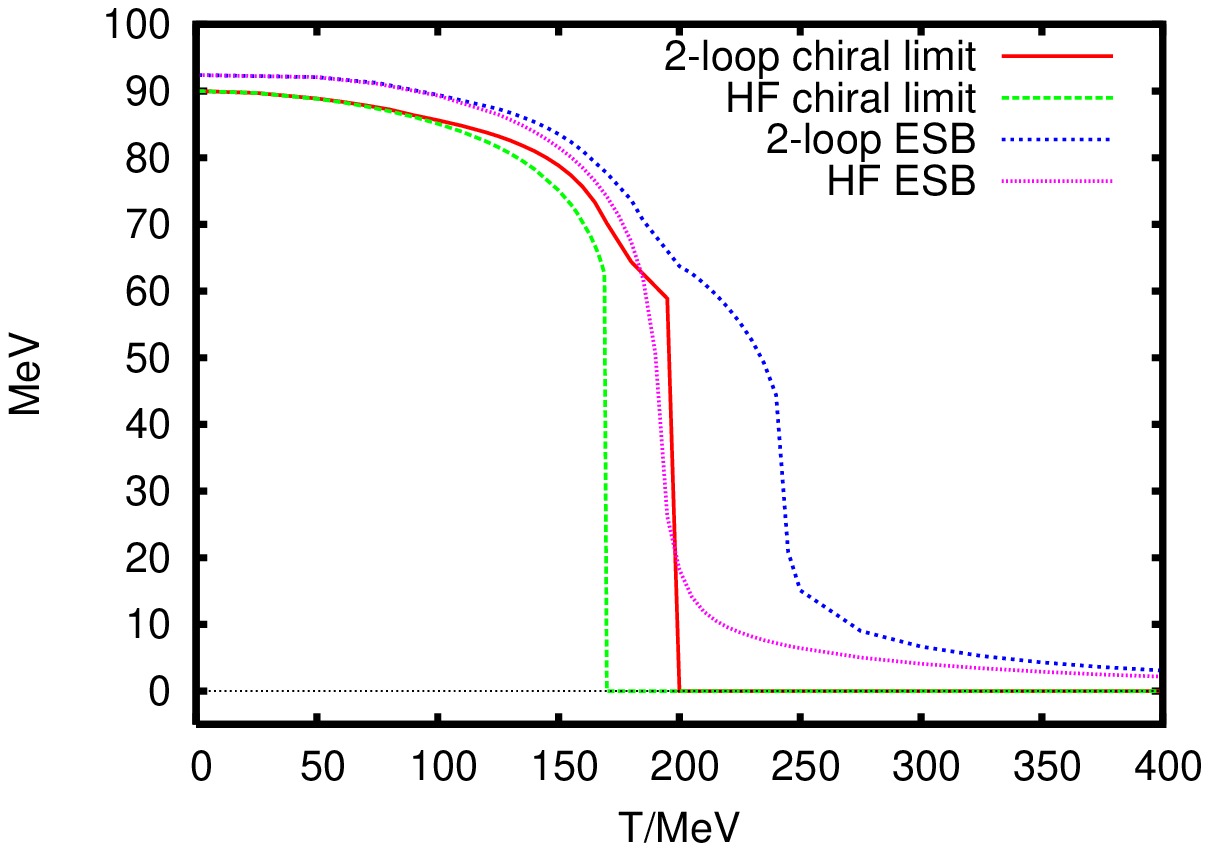}
  \caption{Temperature dependence of the condensate
    in the \Utwo\ model without an axial anomaly. Comparison of the 
    two-loop approximation with \HF\ (HF) for explicit 
    symmetry breaking (ESB)
    and in the chiral limit.}
  \label{fig:u2_noano_condensate}
\end{figure}

For the \Utwo\ model without axial $U(1)$ anomaly we plot
the condensate vs. temperature in Fig.~\ref{fig:u2_noano_condensate}.
In the chiral limit there is a first-order phase transition
at $T_c\approx 200$~MeV ($T_c\approx 170$~MeV in \HF) which
is expected from the respective universality class~\cite{Pisarski:1983ms}.

\begin{figure}
  \centering
  \subfigure[Explicit symmetry breaking.]{
    \includegraphics[width=.8\linewidth]{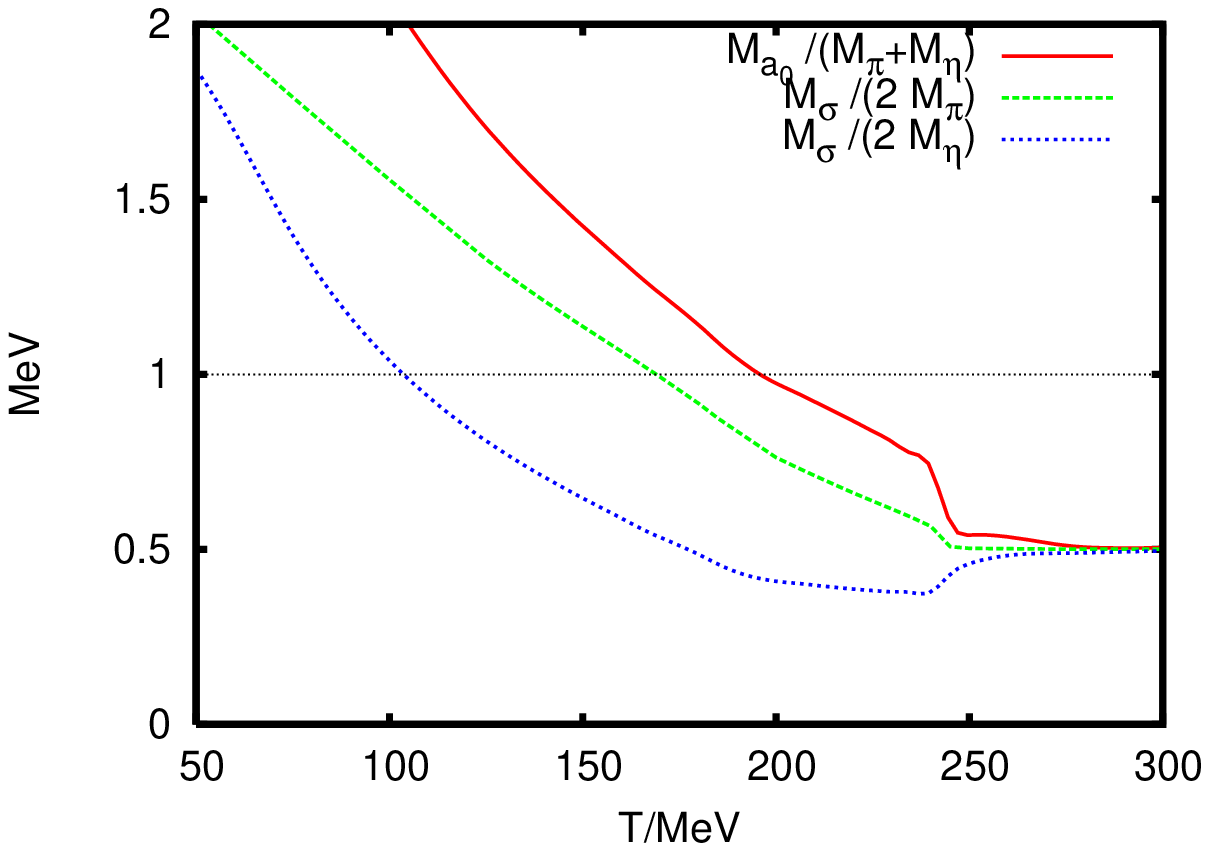}}

  \subfigure[Chiral limit.]{
    \includegraphics[width=.8\linewidth]{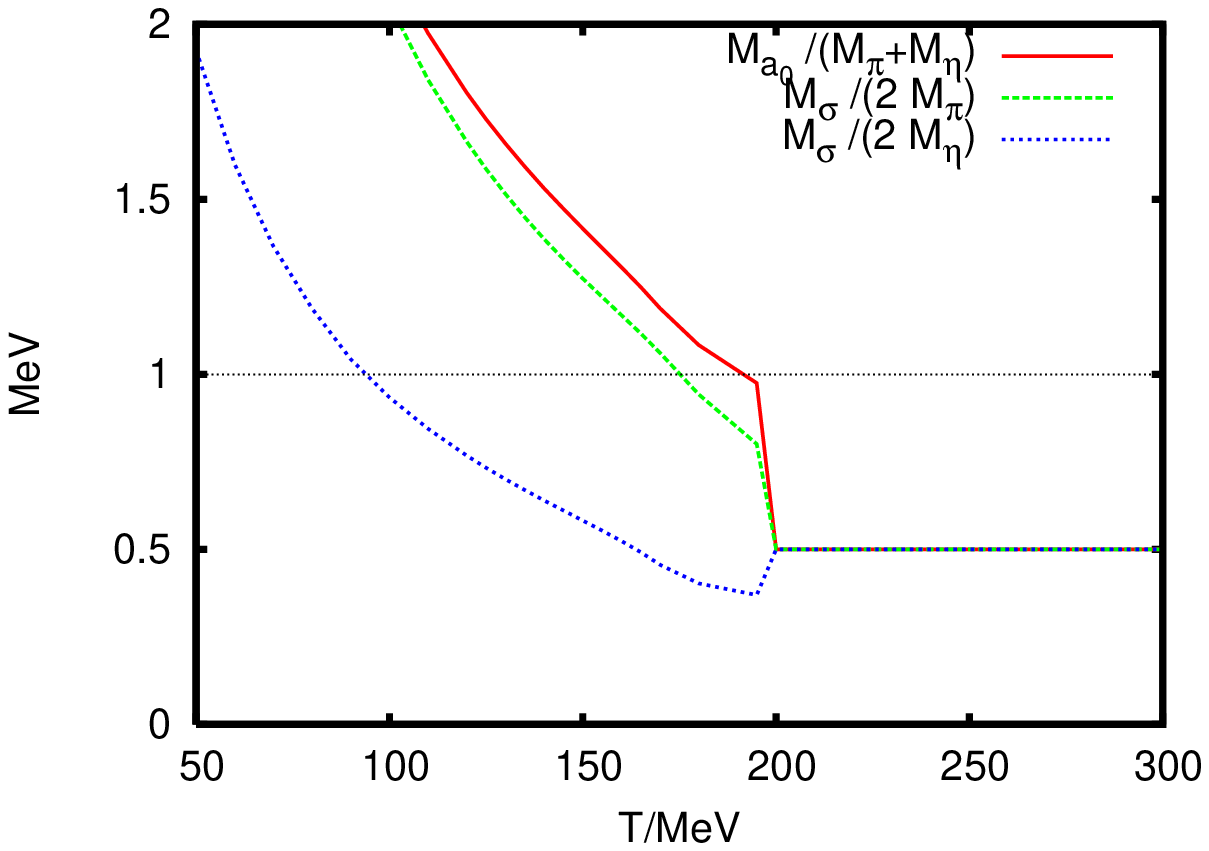}}
  \caption{Thresholds for the decays $\sigma\to \pi\pi$,
    $a_0\to \eta\pi$ and $\sigma\to\eta\eta$
    in the \Utwo\ model without axial anomaly.}
  \label{fig:u2_noano_thresh}
\end{figure}

Here, the deviation from the \HF\ approximation sets 
in at lower temperatures
because the $\sigma\to\eta\eta$ threshold is crossed already at 
temperatures about
100~MeV, followed by $\sigma\to\pi\pi$ at about 175~MeV
and, finally, the on-shell decay $a_0\to\eta\pi$ becomes 
impossible at 200~MeV (see Fig.~\ref{fig:u2_noano_thresh}).
Compared to the case with a finite axial anomaly the effect of the sunset
$S_{\sigma\pi\pi}$ is only about 7\% (instead of 12\%)
of that of the other two. So, in this case as well, the 
deviation from the \HF\ approximation is dominated by
$S_{a_0\eta\pi}$.

\begin{figure}
  \centering
  \subfigure[Explicit symmetry breaking]{
    \includegraphics[width=.8\linewidth]{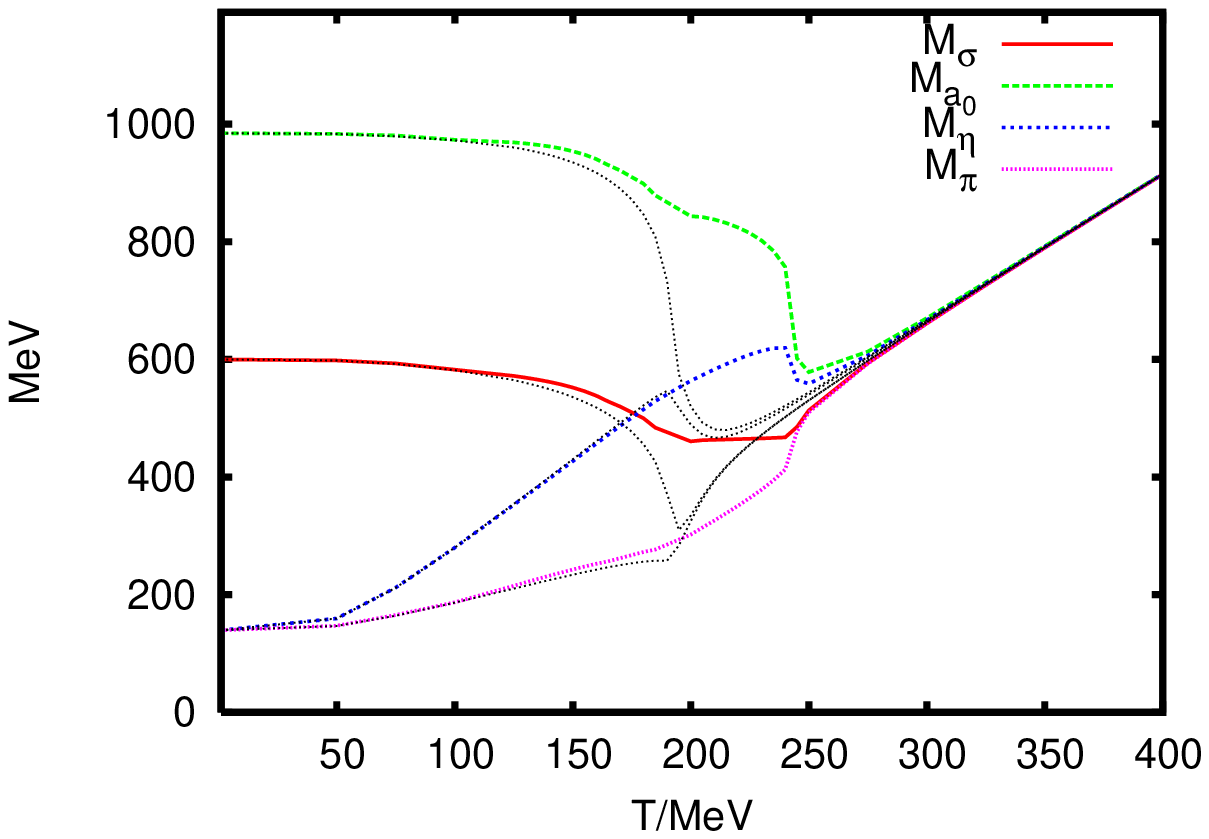}
    \label{fig:masses_u2_esbnoano_600}
  }

  \subfigure[Chiral limit.]{
    \includegraphics[width=.8\linewidth]{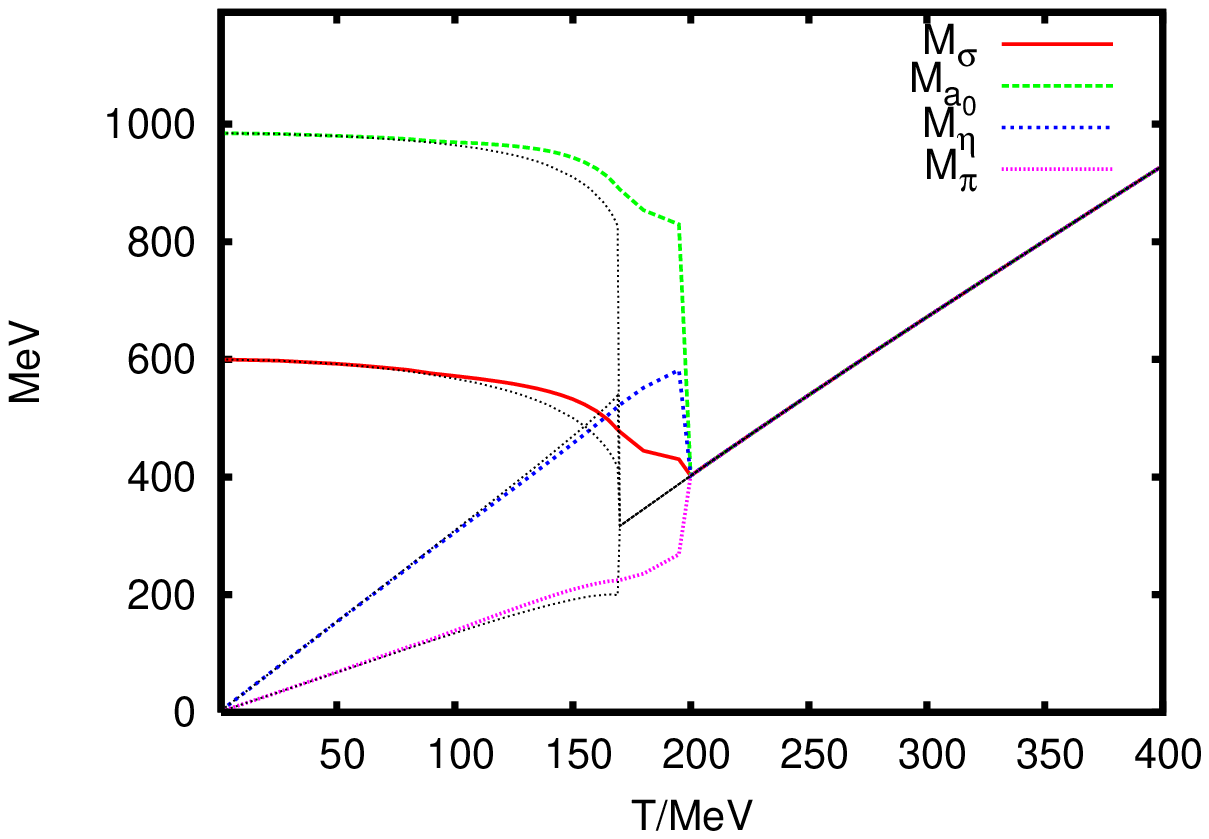}
    \label{fig:u2_ssbnoano_masses}
  }
  \caption{Masses in the \Utwo\ model without axial anomaly 
    as functions of temperature. 
    Comparison of the two-loop approximation
     with \HF\ (dotted lines).}
  \label{fig:u2_noano_masses}
\end{figure}

In Fig.~\ref{fig:masses_u2_esbnoano_600} we observe that, for explicit
symmetry breaking, the masses
of the chiral partners tend to become identical at 250~MeV (200~MeV in \HF)
before the actual restoration of the full \Utwo\
symmetry takes place at about 300~MeV. In the chiral limit,
[Fig.~\ref{fig:u2_ssbnoano_masses}] these
two points coincide when the critical temperature is reached.
At finite temperature the mass of the $\eta$ meson differs
from the pion mass although they both started from 139.6~MeV (or zero in
the chiral limit) at zero 
temperature. This indicates that the approximations considered in
this article
are not well-suited to model the $\eta$ meson as a fourth (pseudo-)Goldstone
boson since they seem to contain an effective $U(1)_A$ breaking through
the unequal treatment of $\eta$ and $\pi$ in the gap 
equations~\eqref{eq:gap_u2}.  This
has been found before in the \HF\ approximation~\cite{Roder:2003uz}.
In the case of a vanishing $U(1)$ anomaly as well, Goldstone's theorem
seems to be violated in the chiral limit as mentioned before because
$M_\eta$ and $M_\pi$ are finite below $T_c$.

\subsubsection{Temperature-dependent $U(1)_A$ anomaly}
\begin{figure}
  \centering
  \includegraphics[width=.8\linewidth]{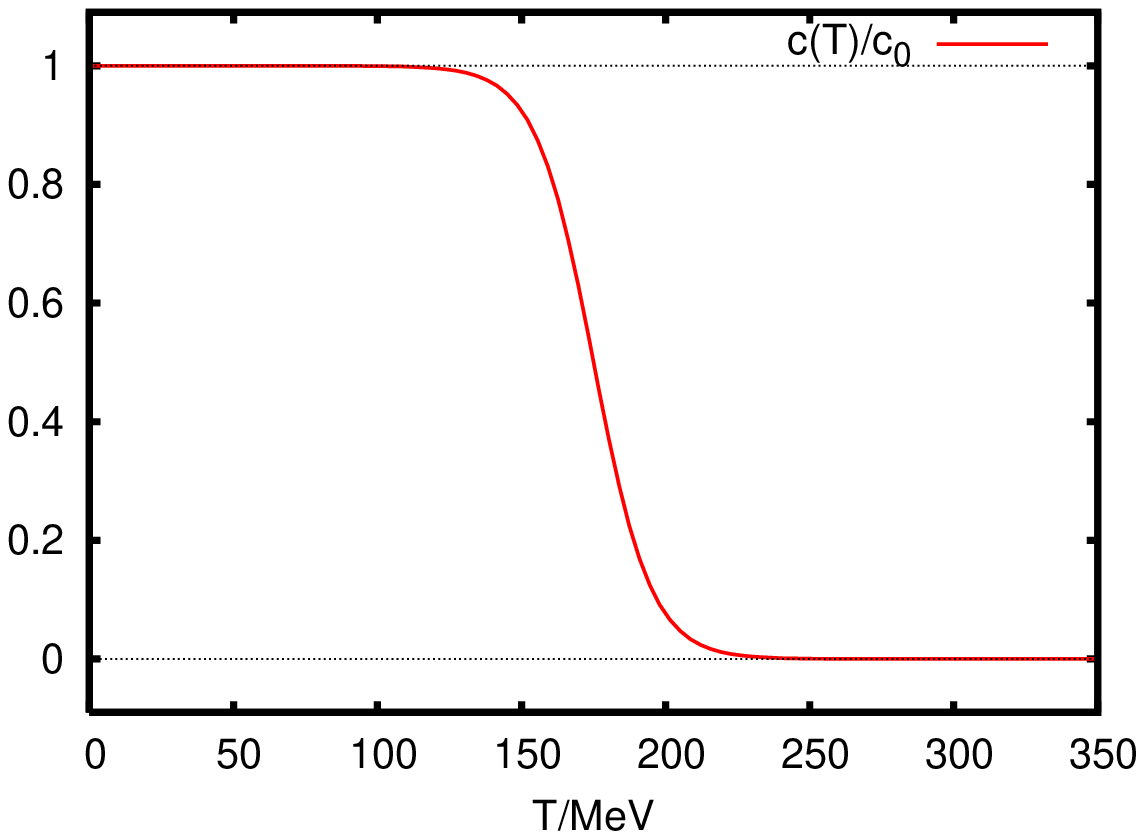}
  \caption{Modelled temperature dependence of the 
    anomaly parameter $c$. The behavior is like the function
    in Eq.~\eqref{eq:cT} with $T_A=175$~MeV and
    $\Delta T=10$~MeV.}
  \label{fig:cT}
\end{figure}

There are indications from the lattice that at high temperatures 
effects arising from the $U(1)_A$ breaking are strongly 
suppressed~\cite{Alles:1996nm,Alles:1998mt,Chu:1994xz,Chu:1997tg,
  Bernard:1996iz,Chandrasekharan:1998yx,Gottlieb:1996ae,
  Kogut:1998rh,Chen:1998xw}. 
This suggests an effective restoration of the $U(1)_A$ symmetry 
close to the critical temperature.
We model this by fixing the parameters at zero temperature
to the physical masses but considering the anomaly parameter $c$
as a function of temperature. As an example we describe a 
suddenly dropping behavior at 175~MeV with the Fermi function
\begin{equation}
  \label{eq:cT}
  c(T)= \frac{c_0}{1+\exp[(T-T_A)/\Delta T]}
\end{equation}
with $T_A=175$~MeV and $\Delta T=10$~MeV, see Fig.~\ref{fig:cT}.
The values are chosen such that at $T\approx 200$~MeV the 
strength of the anomaly has dropped by one order of magnitude.

\begin{figure}
  \centering
  \includegraphics[width=.8\linewidth]{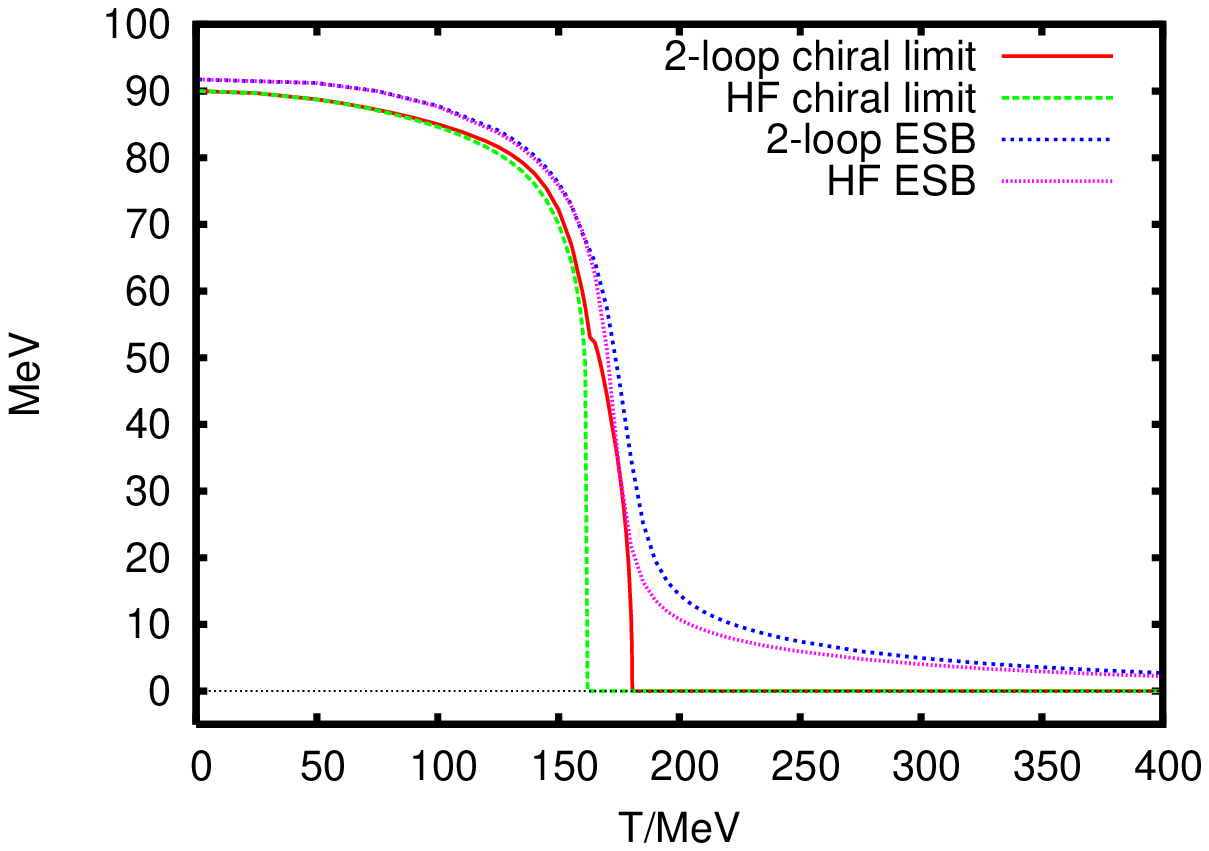}
  \caption{Temperature dependence of the condensate
    in the \Utwo\ model without an axial anomaly. Comparison of the 
    two-loop approximation with \HF\ (HF) for explicit 
    symmetry breaking (ESB) and in the chiral limit.}
  \label{fig:u2_anodyn_condensate}
\end{figure}

\begin{figure}
  \centering
  \subfigure[Explicit symmetry breaking]{
    \includegraphics[width=.8\linewidth]{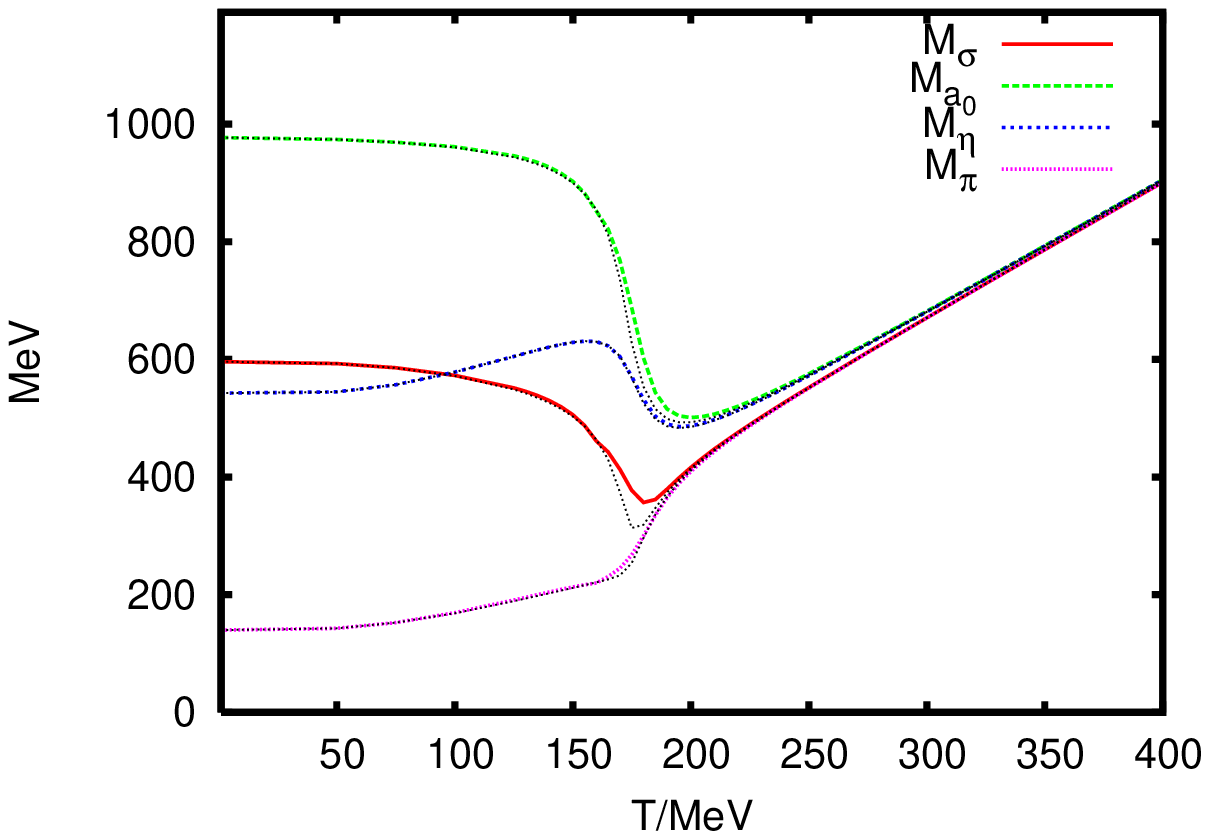}
    \label{fig:u2_esbanodyn_masses}
  }

  \subfigure[Chiral limit.]{
    \includegraphics[width=.8\linewidth]{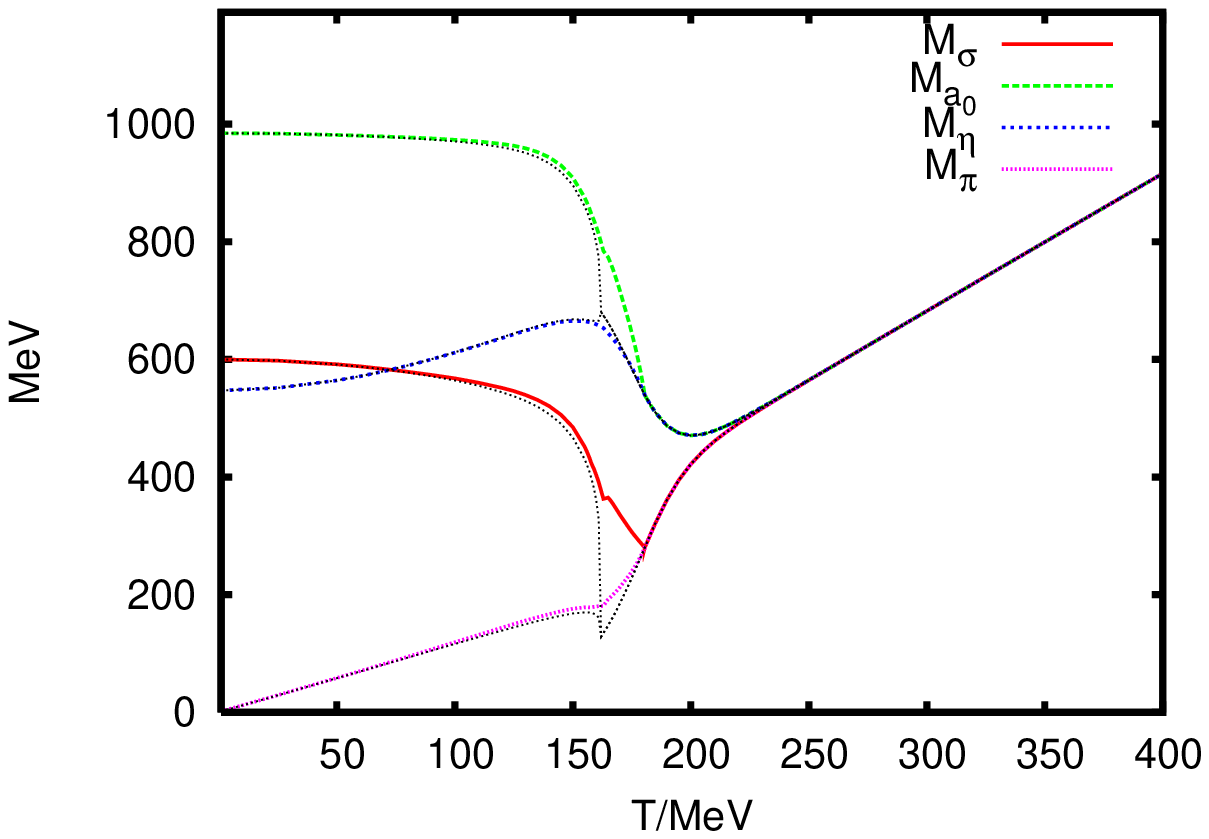}
    \label{fig:u2_ssbanodyn_masses}
  }
  \caption{Masses in the \Utwo\ model with a temperature-dependent 
    axial anomaly as in Fig.~\ref{fig:cT}.
    Comparison of the two-loop approximation
     with \HF\ (dotted lines).}
  \label{fig:u2_anodyn_masses}
\end{figure}

This causes a sharp decrease of the order parameter at about 175~MeV,
cf. Fig.~\ref{fig:u2_anodyn_condensate}. 
In the chiral limit, there is a very weak first-order 
transition at $T_c=180.49$~MeV
where the condensate drops to zero from a value of $1.44$~MeV. In
the \HF\ approximation we find a first-order transition at 
$T\approx 161$~MeV. 
Shifting $T_A$ influences the order of the transition in 
the two-loop approximation; for $T_A\lesssim 175$~MeV
the phase transition is of first order because the anomaly has effectively 
vanished at $T_c$, and for $T_A\gtrsim 175$~MeV there is a second-order
transition because the axial anomaly is sufficiently strong at the
critical temperature. Changing $\Delta T$ in Eq.~\eqref{eq:cT}
has the same effect.
For explicit symmetry breaking, there is a crossover 
at a temperature of about $180$~MeV.
The masses of the chiral
partners become identical at about 200~MeV
(see Fig.~\ref{fig:u2_anodyn_masses}). 
In the two-loop approximation, these temperatures are 
significantly lower than those for the models with a fixed
$U(1)_A$ anomaly. Although the anomaly parameter
tends to zero the full symmetry is only finally restored at about
300~MeV where all four masses become identical. 
The temperature behavior of the condensate and the masses
are quite similar to those found in the framework
of a Nambu--Jona-Lasinio (NJL) model with three
quark flavors~\cite{Costa:2005cz,Costa:2004db}.

The effect that a suddenly dropping anomaly parameter triggers
the restoration of chiral symmetry was observed
earlier in the linear sigma model with three quark 
flavors~\cite{Schaffner-Bielich:1999uj}.
In the chiral limit, the difference between the scalar and the pseudoscalar
meson mass beyond $T_c$ [cf. Fig.~\ref{fig:u2_ssbanodyn_masses}]
is only driven by the function $c(T)$, Eq.~\ref{fig:cT}. So,
chiral symmetry restoration is triggered by the effective
restoration of the $U(1)_A$ symmetry in the chiral limit, too.

\section{Conclusions and outlook}
\label{sec:conclusions-outlook}

We have investigated the restoration of chiral symmetry in the
\Utwo\ and in the $O(N)$ linear
sigma models at finite temperature. 
Taking into account two-loop sunset contributions makes the
condensate drop less rapidly with increasing temperature; 
the crossover and critical temperatures are significantly
higher than in the \HF\ approximation which is 
shown in Table~\ref{tab:temperatures}.
 The deviation from the \HF\ approximation is induced by 
decay threshold effects, especially
of the decay $a_0\to\eta\pi$, which drive the value of the condensate
away from zero.
For the \Utwo\ model with an 
axial $U(1)$ anomaly the masses of the chiral partners
become identical at about 300~MeV for explicit symmetry breaking
or at a critical temperature of about 270~MeV in the chiral limit,
whereas a mass gap of $2c$ between the isospin partners
remains. But even with a zero anomaly parameter, there is an 
effective $U(1)_A$ breaking
by the approximation itself due to an unequal treatment of the $\eta$ 
meson and the pions in the gap equations~\eqref{eq:gap_u2}.
Here as well, the sunsets only raise the temperature where chiral
symmetry is restored compared to the \HF\ approximation.

We have also investigated the effect of a temperature-dependent
anomaly parameter $c(T)$ as in Eq.~\eqref{eq:cT}. 
With a steep decrease at about 175~MeV, such that
the strength is reduced to 10\% at temperatures of about 200~MeV,
we could reduce the transition temperature to 181~MeV in the
chiral limit of the two-loop approximation. 
The phase transition is (very weakly) of first order for this parameters,
though it can be changed to second order by letting $c$ decrease
slower with rising temperature.
The masses of 
the chiral partners become identical
at significantly lower temperatures (about 200~MeV for explicit symmetry
breaking) than with 
a fixed anomaly parameter. Nevertheless, the full symmetry is 
also only restored at about 300~MeV (explicit symmetry breaking) 
or at 220~MeV where
both the condensate and the anomaly have vanished. 
Though it is obvious that a suddenly dropping anomaly parameter 
triggers the restoration of 
chiral symmetry in the two-loop approximation.

There seems to be a violation of Goldstone's theorem in the chiral
limit because all masses of the Goldstone bosons
[$M_\pi$ in Fig.~\ref{fig:u2_ssbano_masses} and
\ref{fig:u2_ssbanodyn_masses}, $M_\pi$ and
$M_\eta$ in Fig.~\ref{fig:u2_ssbnoano_masses}]
are finite below the critical temperature $T_c$.
The reason for that is the fact that resummation schemes
usually violate global symmetries~\cite{vanHees:2002bv}. 
The physical masses, obtained as second derivatives of a
one-particle irreducible effective potential,
are not identical to the variational mass parameters.
The usual geometric argument that proves the physical
Goldstone masses to be zero ist the following.
Consider the second derivative of a 1PI effective potential depending
on the $O(N)$ invariant $\vec{\phi}^2=\phi_a\phi_a$
\begin{equation}
  \frac{\partial^2}{\partial \phi_a \partial \phi_b}
  V(\vec{\phi}^2)\biggr|_{\vec{\phi}=\langle\vec{\phi}\rangle}
  = 2\, V'(\vec{\phi}^2)\, \delta_{ab}
  \biggr|_{\vec{\phi}=\langle\vec{\phi}\rangle}
  + 4\, V''(\vec{\phi}^2)\,\phi_a\phi_b
  \biggr|_{\vec{\phi}=\langle\vec{\phi}\rangle}
  \ .
\end{equation}
With the expectation value pointing only in the $0$-direction,
$\langle\vec{\phi}\rangle=(\phi_0,0,\dots,0)$,
the Goldstone masses are given by the derivatives 
perpendicular to that direction and thus are 
proportional to $V'(\vec{\phi}^2)$
which is zero for a non-trivial vacuum. So, whenever
there is a minimum off zero in the potential there are Goldstone
bosons~\footnote{This argument holds as well for a $U(N_f)\times U(N_f)$
symmetry.}.

Comparing this work with recent 
publications~\cite{Roder:2005vt,Roder:2005qy} one has to state that
the effect of non-local corrections to
the propagators seems to be more efficient for the reduction of the critical
temperature from the \HF\ value of about 200~MeV to a desired value
of about 175~MeV~\cite{Karsch:2001cy,Laermann:2003cv} than considering only
local corrections. The approach described in this article is can only reduce
the crossover temperature significantly if a 
varying anomaly strength is taken into account. 
So, it would be interesting to 
apply non-local resummation schemes also to the \Utwo\ model.

Furthermore, the effective $U(1)_A$ symmetry breaking that is
inherent to the \HF\ and two-loop approximation may be remedied
by a different approximation, possibly by one inspired by a $1/N$ 
expansion similar to that used in Ref.~\cite{Roder:2005qy}.

Including strange mesons ($N_f=3$) could possibly lead to interesting
non-linear effects since the $U(1)_A$ anomaly term is trilinear
for three flavors and thus would generate additional sunset graphs
with different signs. 
Adding fermions (nucleons or constituent quarks) to the model 
using a Yukawa coupling
is also an attractive extension of this work~\cite{michalski06:_param_o_n}. 
For the $O(N)$ model this has been done
by several authors before~\cite{Roh:1996ek,Mocsy:2004ab,Caldas:2000ic}.

 \begin{table}
   \centering
   \begin{tabular}{lllcc}
     \hline\hline
     \multirow{2}*{model} &&& \multicolumn{2}{c}{transition temperature
     (order of phase transition)} \\
     &&& \HF & two-loop \\
     \hline\hline
     \multirow{2}*{$O(4)$}
       && ESB & 225~MeV & 225~MeV\\
       && {chiral limit} 
       & {181 MeV (1st)}& 210~MeV\\
     \hline
     \multirow{4}*{\Utwo} & \multirow{2}*{with anomaly}
       & ESB  & 210~MeV & 270~MeV\\
       && chiral limit & 178 MeV (1st)& 272 MeV (2nd)\\
       & \multirow{2}*{without anomaly}  
       & ESB & 195~MeV & 245~MeV \\
       && chiral limit & 170 MeV (1st)& 200 MeV (1st)\\
       & \multirow{2}*{with varying anomaly}  
       & ESB & 180 MeV & 185 MeV\\
       && chiral limit & 161 MeV (1st)& 181 MeV (weak 1st)\\

     \hline\hline
   \end{tabular}
   \caption{Transition temperatures and order of the
     phase transitions of the two models in two different
     approximations. For explicit symmetry breaking there
     is always a crossover.}
   \label{tab:temperatures}
 \end{table}


\begin{acknowledgments}
  The author was supported by \emph{Deutsche Forschungsgemeinschaft} 
  as a member of \emph{Graduiertenkolleg 841}. Part of this work was
  done during the \emph{Marie Curie Doctoral Training Programme 2005} at ECT*
  in Trento, Italy, which the author enjoyed very much. 
\end{acknowledgments}


\appendix
\renewcommand{\theequation}{\thesection\arabic{equation}}
\setcounter{equation}{0}
\section{Effective action of the \Utwo\ model}
\label{sec:effect-acti-utwo}

This Appendix contains details for the computation of the
effective action of the \Utwo\ model. 
We define the vacuum expectation value
\[ \langle\Phi\rangle = T_a\, \phi_a\]
with a real-valued condensate $\phi_0$ [cf. Eq.~\eqref{eq:vev}]
and shift the (complex) \Utwo\ fields to
\[ \Phi = \frac{1}{2}\phi_0\,\openone + T_a (\sigma_a + i\pi_a)\ ,\]
where $\sigma_a$ and $\pi_a$ are real and symbolize 
the scalar and pseudoscalar meson fields.
The shifted Lagrangian is a sum of four parts,
\begin{equation}
  \label{eq:L_shifted}
  \mathscr{L}[\Phi] = -V_\mathrm{class}(\phi_0) 
  + \mathscr{L}_2[\Phi] + \mathscr{L}_3[\Phi]
  + \mathscr{L}_4[\Phi]
  \ .
\end{equation}
The first part is the classical tree-level potential
\begin{equation}
  \label{eq:L0}
  V_\mathrm{class}(\phi) = \frac{1}{2} m^2 \phi_a^2 - 
  3 \mathcal{G}_{ab}\ \phi_a \phi_b + \frac{1}{3} \mathcal{F}_{abcd}\
  \phi_a \phi_b \phi_c \phi_d - h_a \phi_a\ ,
\end{equation}
which for the choice of Eq.~\eqref{eq:vev} reduces to the expression
in Eq.~\eqref{eq:Vcl}.
The structure of the mass matrices and interactions is given by the
coefficients~\cite{Roder:2003uz}
\begin{subequations}
  \begin{eqnarray}
    \mathcal{G}_{ab} &=& \frac{c}{6} \left(
      \delta_{a0}\delta_{b0} - \delta_{a1}\delta_{b1} 
      - \delta_{a2}\delta_{b2} - \delta_{a3}\delta_{b3} \right)\\
    \mathcal{F}_{abcd} &=& \frac{\lambda_1}{4} \left( 
      \delta_{ab}\delta_{cd} + \delta_{ad}\delta_{bc} 
      + \delta_{ac}\delta_{bd} \right) \nonumber \\
    && + \frac{\lambda_2}{8} \left(
      d_{abn} d_{ncd} + d_{adn} d_{nbc} + d_{acn} d_{nbd} \right)\\
    \mathcal{H}_{abcd} &=& \frac{\lambda_1}{4} \delta_{ab} \delta_{cd}
    + \frac{\lambda_2}{8}\left( d_{abn} d_{ncd} + f_{acn} f_{nbd}
      + f_{bcn} f_{nad} \right)\ .
  \end{eqnarray}
\end{subequations}
The second part of the shifted Lagrangian
consists of all bilinear terms in the fields
\begin{equation}
  \label{eq:L2}
  \begin{split}
    \mathscr{L}_2[\phi_a;\sigma_a,\pi_a] 
    =\ & \frac{1}{2} (\partial_\mu \sigma_a)^2
    - \frac{1}{2} \left[ m^2\ \delta_{ab} - 6\, \mathcal{G}_{ab}
      + 4\, \mathcal{F}_{abcd}\ \phi_c \phi_d \right] \sigma_a \sigma_b\\
    & + \frac{1}{2} (\partial_\mu \pi_a)^2
    - \frac{1}{2} \left[ m^2\ \delta_{ab} + 6\, \mathcal{G}_{ab}
      + 4\, \mathcal{H}_{abcd}\ \phi_c \phi_d \right] \pi_a \pi_b
    \ .
  \end{split}
\end{equation}
Looking at the structure of this equation we conclude that
the tree-level mass matrix is diagonal if the 
expectation value $\phi_a$ has only one
component, \textit{e.g.}, $\phi_a = \delta_{a0}\, \phi_0$. 
So there is neither a  mass mixing among the scalar and pseudoscalar mesons
nor between them. According
to the 2PPI resummation scheme~\cite{Smet:2001un,Verschelde:2000ta,
  Verschelde:1992bs,Baacke:2002pi} one constructs the one-loop
terms as given in Eq.~\eqref{eq:Vq}.

The third part of Eq.~\eqref{eq:L_shifted} contains trilinear terms 
describing the three-particle vertex
\begin{equation}
  \label{eq:L3}
  \begin{split}
    \mathscr{L}_3[\phi_a; \sigma_a, \pi_a] 
    =\ & - \frac{4}{3}\,\mathcal{F}_{abcd}\ \phi_d 
    \sigma_a \sigma_b \sigma_c
    - 4\,\mathcal{H}_{abcd}\ \phi_d \pi_a \pi_b \sigma_c
    \ .
  \end{split}
\end{equation}
The sunset contributions to the effective potential are constructed
from that by computing the expectation value
\begin{equation}
  \label{eq:V_sunset_compute}
  V_\text{sunset}(M^2, \phi_0) = - \frac{1}{2}
  \left< \int_x\int_y
    \mathscr{L}_3(x)\, \mathscr{L}_3(y)
  \right>_\text{2PPI}
\end{equation}
where we chose conventions for Euclidean space-time and 
neglected an overall volume factor on the left hand side.
The subscript 2PPI means that only two-particle point-irreducible
graphs may be constructed from this expectation value resulting
in sunset graphs with three propagators from the vertex at
$x$ to the vertex at $y$ as
required by Wick's theorem. Note that since there is no tree-level
mixing of masses, valid contractions are only of the type
\begin{equation}
  \langle \sigma_a(x)\, \sigma_a(y) \rangle 
  \longrightarrow \frac{1}{p^2+M_{\sigma_a}^2}
  \quad\text{or}\quad
  \langle \pi_a(x)\, \pi_a(y) \rangle
  \longrightarrow \frac{1}{p^2+M_{\pi_a}^2}
\end{equation}
where $\sigma_0=\sigma$, $\sigma_{1,2,3}=a_0$ and $\pi_0=\eta$, 
$\pi_{1,2,3}=\pi$.
This finally yields the sunset contribution to the effective potential
\begin{equation}
  \begin{split}
    V_\mathrm{sunsets}(M^2,\phi_0) = - \phi_0^2\, \Biggl[ & 
    \left( \lambda_1+\frac{3}{2}\lambda_2 \right)^2
    (3\, S_{\sigma\sigma\sigma}+ 3\, S_{\sigma a_0 a_0}
    + S_{\sigma\eta\eta}) \\
    & +  \left( \lambda_1+\frac{\lambda_2}{2} \right)^2
    S_{\sigma\pi\pi} 
    + \frac{3}{2}\lambda_2^2\,S_{a_0\eta\pi}
    \Biggr]\ ,
  \end{split}
\end{equation}
where $S_{ijk}$ denotes a sunset graph with propagators of
the particles $i$, $j$ and $k$.
The last part of the shifted Lagrangian
contains the rest, \textit{i.e.}, the four-vertex
interactions,
\begin{equation}
  \label{eq:L4}
  \begin{split}
    \mathscr{L}_4[\sigma_a, \pi_a] 
    =\ & -2\,\mathcal{H}_{abcd}\ \sigma_a \sigma_b \pi_c \pi_d
    - \frac{1}{3}\,\mathcal{F}_{abcd}\ (\sigma_a\sigma_b\sigma_c\sigma_d
    + \pi_a\pi_b\pi_c\pi_d)\ .
  \end{split}
\end{equation}

We can now write down the general structure of the two-particle
point-irreducible (2PPI) effective potential
\begin{equation}
  \label{eq:V_2PPI}
  V_\mathrm{eff}(M^2,\phi_0) = V_\mathrm{class}(\phi_0)
  + 2\,\mathcal{H}_{abcd}\ \Delta^S_{ab}\, \Delta^P_{cd} 
  + \mathcal{F}_{abcd}\ (\Delta^S_{ab} \Delta^S_{cd}
    + \Delta^P_{ab} \Delta^P_{cd})
    + V_q(M^2, \phi_0)
    \ , 
\end{equation}
where
\begin{equation}
  \label{eq:Delta_ab}
  \Delta^S_{ab} = \langle \sigma_a(x) \sigma_b(x) \rangle
  \quad\text{and}\quad
  \Delta^P_{ab} = \langle \pi_a(x) \pi_b(x) \rangle
\end{equation}
denote expectation values of \emph{local} composite operators 
(``bubbles''), 
and $V_q$ contains higher-order 
corrections made of 2PPI graphs likes sunsets, basketballs
\emph{etc}.
For the masses are diagonal the expectation values~\eqref{eq:Delta_ab}
are diagonal as well. 
Using the physical identification of the meson fields as given in 
Eq.~(\ref{eq:physical_bosons}) we declare
\begin{subequations}
  \begin{align}
    \Delta_\sigma &\equiv \Delta^S_{00} &
    \Delta_\eta &\equiv \Delta^P_{00} \\
    \Delta_{a_0} &\equiv \Delta^S_{11} = \Delta^S_{22} =\Delta^S_{33} &
    \Delta_{\pi} &\equiv \Delta^P_{11} = \Delta^P_{22} =\Delta^P_{33} &
    \ .
  \end{align}
\end{subequations}
Contracting the coefficients $\mathcal{F}_{abcd}$ and $\mathcal{H}_{abcd}$
with the appropriate expectation values we obtain the following expression 
for the 2PPI effective action
\begin{equation}
  \label{eq:V_db_app}
  \begin{split}
    V_\mathrm{eff} (M^2,\phi_0)
    =\ & V_\mathrm{class}(\phi_0) + V_q(\phi_0,M^2) \\
    & - \left(\frac{\lambda_1}{4}+ \frac{\lambda_2}{8}\right)
    (3\, \Delta_\sigma^2 + 15\,\Delta_\pi^2 + 6\, \Delta_\sigma\Delta_\pi) \\
    & - \left(\frac{\lambda_1}{4}+ \frac{\lambda_2}{8}\right)
    (3\, \Delta_\eta^2 + 15\,\Delta_{a_0}^2 + 6\, \Delta_\eta\Delta_{a_0}) \\
    & - \left( \frac{\lambda_1}{2}+\frac{\lambda_2}{4}\right)
    \Delta_\sigma\Delta_\eta 
    + 3\,\left( \frac{\lambda_1}{2}+\frac{3}{4}\lambda_2\right)
    \Delta_{a_0}\Delta_\sigma \\
    & - 3\,\left( \frac{\lambda_1}{2}+\frac{3}{4}\lambda_2\right)
    \Delta_\pi\Delta_\eta 
    + 3\,\left( \frac{3}{2}\lambda_1+\frac{7}{4}\lambda_2\right)
    \Delta_{a_0}\Delta_\pi
    \ ,
  \end{split}
\end{equation}
where $V_\mathrm{eff}$ is a function of the condensate $\phi_0$ and
all four masses $M_\sigma$, $M_{a_0}$, $M_\eta$ and $M_\pi$. 
Note that this is the case for each quantity $\Delta_*$ as well
since it is given by Eq.~(\ref{eq:Delta}).
Equation~\eqref{eq:V_db_app} looks like the double-bubble term
in the 2PI expansion~\cite{Roder:2003uz} except for the sign
which is different here due to the special role of this contribution
and the quantity $\Delta$ within the 2PPI 
formalism~\cite{Smet:2001un,Verschelde:2000ta,Verschelde:1992bs}.

For a detailed graphical comparison of the 2PPI and the 2PI 
effective action see the Appendix of Ref.~\cite{Baacke:2002pi}.


\bibliography{sigma2prd}

\begin{thebibliography}{60}
\expandafter\ifx\csname natexlab\endcsname\relax\def\natexlab#1{#1}\fi
\expandafter\ifx\csname bibnamefont\endcsname\relax
  \def\bibnamefont#1{#1}\fi
\expandafter\ifx\csname bibfnamefont\endcsname\relax
  \def\bibfnamefont#1{#1}\fi
\expandafter\ifx\csname citenamefont\endcsname\relax
  \def\citenamefont#1{#1}\fi
\expandafter\ifx\csname url\endcsname\relax
  \def\url#1{\texttt{#1}}\fi
\expandafter\ifx\csname urlprefix\endcsname\relax\def\urlprefix{URL }\fi
\providecommand{\bibinfo}[2]{#2}
\providecommand{\eprint}[2][]{\url{#2}}

\bibitem[{\citenamefont{'t~Hooft}(1976{\natexlab{a}})}]{'tHooft:1976up}
\bibinfo{author}{\bibfnamefont{G.}~\bibnamefont{'t~Hooft}},
  \bibinfo{journal}{Phys. Rev. Lett.} \textbf{\bibinfo{volume}{37}},
  \bibinfo{pages}{8} (\bibinfo{year}{1976}{\natexlab{a}}).

\bibitem[{\citenamefont{'t~Hooft}(1976{\natexlab{b}})}]{'tHooft:1976fv}
\bibinfo{author}{\bibfnamefont{G.}~\bibnamefont{'t~Hooft}},
  \bibinfo{journal}{Phys. Rev.} \textbf{\bibinfo{volume}{D14}},
  \bibinfo{pages}{3432} (\bibinfo{year}{1976}{\natexlab{b}}).

\bibitem[{\citenamefont{Pisarski and Wilczek}(1984)}]{Pisarski:1983ms}
\bibinfo{author}{\bibfnamefont{R.~D.} \bibnamefont{Pisarski}} \bibnamefont{and}
  \bibinfo{author}{\bibfnamefont{F.}~\bibnamefont{Wilczek}},
  \bibinfo{journal}{Phys. Rev.} \textbf{\bibinfo{volume}{D29}},
  \bibinfo{pages}{338} (\bibinfo{year}{1984}).

\bibitem[{\citenamefont{Karsch}(2002)}]{Karsch:2001cy}
\bibinfo{author}{\bibfnamefont{F.}~\bibnamefont{Karsch}},
  \bibinfo{journal}{Lect. Notes Phys.} \textbf{\bibinfo{volume}{583}},
  \bibinfo{pages}{209} (\bibinfo{year}{2002}), \eprint{hep-lat/0106019}.

\bibitem[{\citenamefont{Laermann and Philipsen}(2003)}]{Laermann:2003cv}
\bibinfo{author}{\bibfnamefont{E.}~\bibnamefont{Laermann}} \bibnamefont{and}
  \bibinfo{author}{\bibfnamefont{O.}~\bibnamefont{Philipsen}},
  \bibinfo{journal}{Ann. Rev. Nucl. Part. Sci.} \textbf{\bibinfo{volume}{53}},
  \bibinfo{pages}{163} (\bibinfo{year}{2003}), \eprint{hep-ph/0303042}.

\bibitem[{\citenamefont{Fodor and Katz}(2004)}]{Fodor:2004nz}
\bibinfo{author}{\bibfnamefont{Z.}~\bibnamefont{Fodor}} \bibnamefont{and}
  \bibinfo{author}{\bibfnamefont{S.~D.} \bibnamefont{Katz}},
  \bibinfo{journal}{JHEP} \textbf{\bibinfo{volume}{04}}, \bibinfo{pages}{050}
  (\bibinfo{year}{2004}), \eprint{hep-lat/0402006}.

\bibitem[{\citenamefont{de~Forcrand and Philipsen}(2002)}]{deForcrand:2002ci}
\bibinfo{author}{\bibfnamefont{P.}~\bibnamefont{de~Forcrand}} \bibnamefont{and}
  \bibinfo{author}{\bibfnamefont{O.}~\bibnamefont{Philipsen}},
  \bibinfo{journal}{Nucl. Phys.} \textbf{\bibinfo{volume}{B642}},
  \bibinfo{pages}{290} (\bibinfo{year}{2002}), \eprint{hep-lat/0205016}.

\bibitem[{\citenamefont{Gell-Mann and Levy}(1960)}]{Gell-Mann:1960np}
\bibinfo{author}{\bibfnamefont{M.}~\bibnamefont{Gell-Mann}} \bibnamefont{and}
  \bibinfo{author}{\bibfnamefont{M.}~\bibnamefont{Levy}},
  \bibinfo{journal}{Nuovo Cim.} \textbf{\bibinfo{volume}{16}},
  \bibinfo{pages}{705} (\bibinfo{year}{1960}).

\bibitem[{\citenamefont{Dolan and Jackiw}(1974)}]{Dolan:1973qd}
\bibinfo{author}{\bibfnamefont{L.}~\bibnamefont{Dolan}} \bibnamefont{and}
  \bibinfo{author}{\bibfnamefont{R.}~\bibnamefont{Jackiw}},
  \bibinfo{journal}{Phys. Rev.} \textbf{\bibinfo{volume}{D9}},
  \bibinfo{pages}{3320} (\bibinfo{year}{1974}).

\bibitem[{\citenamefont{Braaten and
  Pisarski}(1990{\natexlab{a}})}]{Braaten:1989kk}
\bibinfo{author}{\bibfnamefont{E.}~\bibnamefont{Braaten}} \bibnamefont{and}
  \bibinfo{author}{\bibfnamefont{R.~D.} \bibnamefont{Pisarski}},
  \bibinfo{journal}{Phys. Rev. Lett.} \textbf{\bibinfo{volume}{64}},
  \bibinfo{pages}{1338} (\bibinfo{year}{1990}{\natexlab{a}}).

\bibitem[{\citenamefont{Braaten and
  Pisarski}(1990{\natexlab{b}})}]{Braaten:1989mz}
\bibinfo{author}{\bibfnamefont{E.}~\bibnamefont{Braaten}} \bibnamefont{and}
  \bibinfo{author}{\bibfnamefont{R.~D.} \bibnamefont{Pisarski}},
  \bibinfo{journal}{Nucl. Phys.} \textbf{\bibinfo{volume}{B337}},
  \bibinfo{pages}{569} (\bibinfo{year}{1990}{\natexlab{b}}).

\bibitem[{\citenamefont{Parwani}(1992)}]{Parwani:1991gq}
\bibinfo{author}{\bibfnamefont{R.~R.} \bibnamefont{Parwani}},
  \bibinfo{journal}{Phys. Rev.} \textbf{\bibinfo{volume}{D45}},
  \bibinfo{pages}{4695} (\bibinfo{year}{1992}), \eprint{hep-ph/9204216}.

\bibitem[{\citenamefont{Verschelde and Coppens}(1992)}]{Verschelde:1992bs}
\bibinfo{author}{\bibfnamefont{H.}~\bibnamefont{Verschelde}} \bibnamefont{and}
  \bibinfo{author}{\bibfnamefont{M.}~\bibnamefont{Coppens}},
  \bibinfo{journal}{Phys. Lett.} \textbf{\bibinfo{volume}{B287}},
  \bibinfo{pages}{133} (\bibinfo{year}{1992}).

\bibitem[{\citenamefont{Cornwall et~al.}(1974)\citenamefont{Cornwall, Jackiw,
  and Tomboulis}}]{Cornwall:1974vz}
\bibinfo{author}{\bibfnamefont{J.~M.} \bibnamefont{Cornwall}},
  \bibinfo{author}{\bibfnamefont{R.}~\bibnamefont{Jackiw}}, \bibnamefont{and}
  \bibinfo{author}{\bibfnamefont{E.}~\bibnamefont{Tomboulis}},
  \bibinfo{journal}{Phys. Rev.} \textbf{\bibinfo{volume}{D10}},
  \bibinfo{pages}{2428} (\bibinfo{year}{1974}).

\bibitem[{\citenamefont{{R\"oder} et~al.}(2003)\citenamefont{{R\"oder},
  Ruppert, and Rischke}}]{Roder:2003uz}
\bibinfo{author}{\bibfnamefont{D.}~\bibnamefont{{R\"oder}}},
  \bibinfo{author}{\bibfnamefont{J.}~\bibnamefont{Ruppert}}, \bibnamefont{and}
  \bibinfo{author}{\bibfnamefont{D.~H.} \bibnamefont{Rischke}},
  \bibinfo{journal}{Phys. Rev.} \textbf{\bibinfo{volume}{D68}},
  \bibinfo{pages}{016003} (\bibinfo{year}{2003}), \eprint{nucl-th/0301085}.

\bibitem[{\citenamefont{Lenaghan et~al.}(2000)\citenamefont{Lenaghan, Rischke,
  and Schaffner-Bielich}}]{Lenaghan:2000ey}
\bibinfo{author}{\bibfnamefont{J.~T.} \bibnamefont{Lenaghan}},
  \bibinfo{author}{\bibfnamefont{D.~H.} \bibnamefont{Rischke}},
  \bibnamefont{and}
  \bibinfo{author}{\bibfnamefont{J.}~\bibnamefont{Schaffner-Bielich}},
  \bibinfo{journal}{Phys. Rev.} \textbf{\bibinfo{volume}{D62}},
  \bibinfo{pages}{085008} (\bibinfo{year}{2000}), \eprint{nucl-th/0004006}.

\bibitem[{\citenamefont{Schaffner-Bielich}(2000)}]{Schaffner-Bielich:1999uj}
\bibinfo{author}{\bibfnamefont{J.}~\bibnamefont{Schaffner-Bielich}},
  \bibinfo{journal}{Phys. Rev. Lett.} \textbf{\bibinfo{volume}{84}},
  \bibinfo{pages}{3261} (\bibinfo{year}{2000}), \eprint{hep-ph/9906361}.

\bibitem[{\citenamefont{Geddes}(1980)}]{Geddes:1979nd}
\bibinfo{author}{\bibfnamefont{H.~B.} \bibnamefont{Geddes}},
  \bibinfo{journal}{Phys. Rev.} \textbf{\bibinfo{volume}{D21}},
  \bibinfo{pages}{278} (\bibinfo{year}{1980}).

\bibitem[{\citenamefont{Lenaghan and Rischke}(2000)}]{Lenaghan:1999si}
\bibinfo{author}{\bibfnamefont{J.~T.} \bibnamefont{Lenaghan}} \bibnamefont{and}
  \bibinfo{author}{\bibfnamefont{D.~H.} \bibnamefont{Rischke}},
  \bibinfo{journal}{J. Phys.} \textbf{\bibinfo{volume}{G26}},
  \bibinfo{pages}{431} (\bibinfo{year}{2000}), \eprint{nucl-th/9901049}.

\bibitem[{\citenamefont{Baacke and
  Michalski}(2003{\natexlab{a}})}]{Baacke:2002pi}
\bibinfo{author}{\bibfnamefont{J.}~\bibnamefont{Baacke}} \bibnamefont{and}
  \bibinfo{author}{\bibfnamefont{S.}~\bibnamefont{Michalski}},
  \bibinfo{journal}{Phys. Rev.} \textbf{\bibinfo{volume}{D67}},
  \bibinfo{pages}{085006} (\bibinfo{year}{2003}{\natexlab{a}}),
  \eprint{hep-ph/0210060}.

\bibitem[{\citenamefont{Petropoulos}(1999)}]{Petropoulos:1998gt}
\bibinfo{author}{\bibfnamefont{N.}~\bibnamefont{Petropoulos}},
  \bibinfo{journal}{J. Phys.} \textbf{\bibinfo{volume}{G25}},
  \bibinfo{pages}{2225} (\bibinfo{year}{1999}), \eprint{hep-ph/9807331}.

\bibitem[{\citenamefont{Patk{\'o}s et~al.}(2002)\citenamefont{Patk{\'o}s,
  Sz{\'e}p, and Sz{\'e}pfalusy}}]{Patkos:2002xb}
\bibinfo{author}{\bibfnamefont{A.}~\bibnamefont{Patk{\'o}s}},
  \bibinfo{author}{\bibfnamefont{Z.}~\bibnamefont{Sz{\'e}p}}, \bibnamefont{and}
  \bibinfo{author}{\bibfnamefont{P.}~\bibnamefont{Sz{\'e}pfalusy}},
  \bibinfo{journal}{Phys. Lett.} \textbf{\bibinfo{volume}{B537}},
  \bibinfo{pages}{77} (\bibinfo{year}{2002}), \eprint{hep-ph/0202261}.

\bibitem[{\citenamefont{Chiku and Hatsuda}(1998)}]{Chiku:1998kd}
\bibinfo{author}{\bibfnamefont{S.}~\bibnamefont{Chiku}} \bibnamefont{and}
  \bibinfo{author}{\bibfnamefont{T.}~\bibnamefont{Hatsuda}},
  \bibinfo{journal}{Phys. Rev.} \textbf{\bibinfo{volume}{D58}},
  \bibinfo{pages}{076001} (\bibinfo{year}{1998}), \eprint{hep-ph/9803226}.

\bibitem[{\citenamefont{Nemoto et~al.}(2000)\citenamefont{Nemoto, Naito, and
  Oka}}]{Nemoto:1999qf}
\bibinfo{author}{\bibfnamefont{Y.}~\bibnamefont{Nemoto}},
  \bibinfo{author}{\bibfnamefont{K.}~\bibnamefont{Naito}}, \bibnamefont{and}
  \bibinfo{author}{\bibfnamefont{M.}~\bibnamefont{Oka}}, \bibinfo{journal}{Eur.
  Phys. J.} \textbf{\bibinfo{volume}{A9}}, \bibinfo{pages}{245}
  (\bibinfo{year}{2000}), \eprint{hep-ph/9911431}.

\bibitem[{\citenamefont{Smet et~al.}(2002)\citenamefont{Smet, Vanzielighem, van
  Acoleyen, and Verschelde}}]{Smet:2001un}
\bibinfo{author}{\bibfnamefont{G.}~\bibnamefont{Smet}},
  \bibinfo{author}{\bibfnamefont{T.}~\bibnamefont{Vanzielighem}},
  \bibinfo{author}{\bibfnamefont{K.}~\bibnamefont{van Acoleyen}},
  \bibnamefont{and}
  \bibinfo{author}{\bibfnamefont{H.}~\bibnamefont{Verschelde}},
  \bibinfo{journal}{Phys. Rev.} \textbf{\bibinfo{volume}{D65}},
  \bibinfo{pages}{045015} (\bibinfo{year}{2002}), \eprint{hep-th/0108163}.

\bibitem[{\citenamefont{Verschelde and De~Pessemier}(2002)}]{Verschelde:2000ta}
\bibinfo{author}{\bibfnamefont{H.}~\bibnamefont{Verschelde}} \bibnamefont{and}
  \bibinfo{author}{\bibfnamefont{J.}~\bibnamefont{De~Pessemier}},
  \bibinfo{journal}{Eur. Phys. J.} \textbf{\bibinfo{volume}{C22}},
  \bibinfo{pages}{771} (\bibinfo{year}{2002}), \eprint{hep-th/0009241}.

\bibitem[{\citenamefont{R{\"o}der et~al.}(2005)\citenamefont{R{\"o}der,
  Ruppert, and Rischke}}]{Roder:2005vt}
\bibinfo{author}{\bibfnamefont{D.}~\bibnamefont{R{\"o}der}},
  \bibinfo{author}{\bibfnamefont{J.}~\bibnamefont{Ruppert}}, \bibnamefont{and}
  \bibinfo{author}{\bibfnamefont{D.~H.} \bibnamefont{Rischke}}
  (\bibinfo{year}{2005}), \eprint{hep-ph/0503042}.

\bibitem[{\citenamefont{R{\"o}der}(2005)}]{Roder:2005qy}
\bibinfo{author}{\bibfnamefont{D.}~\bibnamefont{R{\"o}der}}
  (\bibinfo{year}{2005}), \eprint{hep-ph/0509232}.

\bibitem[{\citenamefont{Baacke and
  Michalski}(2004{\natexlab{a}})}]{Baacke:2004xm}
\bibinfo{author}{\bibfnamefont{J.}~\bibnamefont{Baacke}} \bibnamefont{and}
  \bibinfo{author}{\bibfnamefont{S.}~\bibnamefont{Michalski}}
  (\bibinfo{year}{2004}{\natexlab{a}}), \eprint{hep-ph/0409153}.

\bibitem[{\citenamefont{Baacke and
  Michalski}(2004{\natexlab{b}})}]{Baacke:2004dp}
\bibinfo{author}{\bibfnamefont{J.}~\bibnamefont{Baacke}} \bibnamefont{and}
  \bibinfo{author}{\bibfnamefont{S.}~\bibnamefont{Michalski}},
  \bibinfo{journal}{Phys. Rev.} \textbf{\bibinfo{volume}{D70}},
  \bibinfo{pages}{085002} (\bibinfo{year}{2004}{\natexlab{b}}),
  \eprint{hep-ph/0407152}.

\bibitem[{\citenamefont{Andersen et~al.}(2004)\citenamefont{Andersen, Boer, and
  Warringa}}]{Andersen:2004ae}
\bibinfo{author}{\bibfnamefont{J.~O.} \bibnamefont{Andersen}},
  \bibinfo{author}{\bibfnamefont{D.}~\bibnamefont{Boer}}, \bibnamefont{and}
  \bibinfo{author}{\bibfnamefont{H.~J.} \bibnamefont{Warringa}},
  \bibinfo{journal}{Phys. Rev.} \textbf{\bibinfo{volume}{D70}},
  \bibinfo{pages}{116007} (\bibinfo{year}{2004}), \eprint{hep-ph/0408033}.

\bibitem[{\citenamefont{de~Godoy~Caldas}(2002)}]{deGodoyCaldas:2001mb}
\bibinfo{author}{\bibfnamefont{H.~C.} \bibnamefont{de~Godoy~Caldas}},
  \bibinfo{journal}{Phys. Rev.} \textbf{\bibinfo{volume}{D65}},
  \bibinfo{pages}{065005} (\bibinfo{year}{2002}), \eprint{hep-th/0111194}.

\bibitem[{\citenamefont{van Hees and
  Knoll}(2002{\natexlab{a}})}]{vanHees:2001ik}
\bibinfo{author}{\bibfnamefont{H.}~\bibnamefont{van Hees}} \bibnamefont{and}
  \bibinfo{author}{\bibfnamefont{J.}~\bibnamefont{Knoll}},
  \bibinfo{journal}{Phys. Rev.} \textbf{\bibinfo{volume}{D65}},
  \bibinfo{pages}{025010} (\bibinfo{year}{2002}{\natexlab{a}}),
  \eprint{hep-ph/0107200}.

\bibitem[{\citenamefont{van Hees and
  Knoll}(2002{\natexlab{b}})}]{VanHees:2001pf}
\bibinfo{author}{\bibfnamefont{H.}~\bibnamefont{van Hees}} \bibnamefont{and}
  \bibinfo{author}{\bibfnamefont{J.}~\bibnamefont{Knoll}},
  \bibinfo{journal}{Phys. Rev.} \textbf{\bibinfo{volume}{D65}},
  \bibinfo{pages}{105005} (\bibinfo{year}{2002}{\natexlab{b}}),
  \eprint{hep-ph/0111193}.

\bibitem[{\citenamefont{van Hees and
  Knoll}(2002{\natexlab{c}})}]{vanHees:2002bv}
\bibinfo{author}{\bibfnamefont{H.}~\bibnamefont{van Hees}} \bibnamefont{and}
  \bibinfo{author}{\bibfnamefont{J.}~\bibnamefont{Knoll}},
  \bibinfo{journal}{Phys. Rev.} \textbf{\bibinfo{volume}{D66}},
  \bibinfo{pages}{025028} (\bibinfo{year}{2002}{\natexlab{c}}),
  \eprint{hep-ph/0203008}.

\bibitem[{\citenamefont{Berges et~al.}(2005)\citenamefont{Berges, Bors{\'a}nyi,
  Reinosa, and Serreau}}]{Berges:2004hn}
\bibinfo{author}{\bibfnamefont{J.}~\bibnamefont{Berges}},
  \bibinfo{author}{\bibfnamefont{S.}~\bibnamefont{Bors{\'a}nyi}},
  \bibinfo{author}{\bibfnamefont{U.}~\bibnamefont{Reinosa}}, \bibnamefont{and}
  \bibinfo{author}{\bibfnamefont{J.}~\bibnamefont{Serreau}},
  \bibinfo{journal}{Phys. Rev.} \textbf{\bibinfo{volume}{D71}},
  \bibinfo{pages}{105004} (\bibinfo{year}{2005}), \eprint{hep-ph/0409123}.

\bibitem[{\citenamefont{Blaizot et~al.}(2004)\citenamefont{Blaizot, Iancu, and
  Reinosa}}]{Blaizot:2003an}
\bibinfo{author}{\bibfnamefont{J.-P.} \bibnamefont{Blaizot}},
  \bibinfo{author}{\bibfnamefont{E.}~\bibnamefont{Iancu}}, \bibnamefont{and}
  \bibinfo{author}{\bibfnamefont{U.}~\bibnamefont{Reinosa}},
  \bibinfo{journal}{Nucl. Phys.} \textbf{\bibinfo{volume}{A736}},
  \bibinfo{pages}{149} (\bibinfo{year}{2004}), \eprint{hep-ph/0312085}.

\bibitem[{\citenamefont{Ivanov et~al.}(2005)\citenamefont{Ivanov, Riek, van
  Hees, and Knoll}}]{Ivanov:2005bv}
\bibinfo{author}{\bibfnamefont{Y.~B.} \bibnamefont{Ivanov}},
  \bibinfo{author}{\bibfnamefont{F.}~\bibnamefont{Riek}},
  \bibinfo{author}{\bibfnamefont{H.}~\bibnamefont{van Hees}}, \bibnamefont{and}
  \bibinfo{author}{\bibfnamefont{J.}~\bibnamefont{Knoll}},
  \bibinfo{journal}{Phys. Rev.} \textbf{\bibinfo{volume}{D72}},
  \bibinfo{pages}{036008} (\bibinfo{year}{2005}), \eprint{hep-ph/0506157}.

\bibitem[{\citenamefont{Destri and
  Sartirana}(2005{\natexlab{a}})}]{Destri:2005se}
\bibinfo{author}{\bibfnamefont{C.}~\bibnamefont{Destri}} \bibnamefont{and}
  \bibinfo{author}{\bibfnamefont{A.}~\bibnamefont{Sartirana}}
  (\bibinfo{year}{2005}{\natexlab{a}}), \eprint{hep-ph/0509032}.

\bibitem[{\citenamefont{Destri and
  Sartirana}(2005{\natexlab{b}})}]{Destri:2005qm}
\bibinfo{author}{\bibfnamefont{C.}~\bibnamefont{Destri}} \bibnamefont{and}
  \bibinfo{author}{\bibfnamefont{A.}~\bibnamefont{Sartirana}},
  \bibinfo{journal}{Phys. Rev.} \textbf{\bibinfo{volume}{D72}},
  \bibinfo{pages}{065003} (\bibinfo{year}{2005}{\natexlab{b}}),
  \eprint{hep-ph/0504029}.

\bibitem[{\citenamefont{Vafa and Witten}(1984)}]{Vafa:1983tf}
\bibinfo{author}{\bibfnamefont{C.}~\bibnamefont{Vafa}} \bibnamefont{and}
  \bibinfo{author}{\bibfnamefont{E.}~\bibnamefont{Witten}},
  \bibinfo{journal}{Nucl. Phys.} \textbf{\bibinfo{volume}{B234}},
  \bibinfo{pages}{173} (\bibinfo{year}{1984}).

\bibitem[{\citenamefont{Baacke and
  Michalski}(2003{\natexlab{b}})}]{Baacke:2003dk}
\bibinfo{author}{\bibfnamefont{J.}~\bibnamefont{Baacke}} \bibnamefont{and}
  \bibinfo{author}{\bibfnamefont{S.}~\bibnamefont{Michalski}}
  (\bibinfo{year}{2003}{\natexlab{b}}), \eprint{hep-ph/0312031}.

\bibitem[{\citenamefont{Eidelman et~al.}(2004)}]{Eidelman:2004wy}
\bibinfo{author}{\bibfnamefont{S.}~\bibnamefont{Eidelman}} \bibnamefont{et~al.}
  (\bibinfo{collaboration}{Particle Data Group}), \bibinfo{journal}{Phys.
  Lett.} \textbf{\bibinfo{volume}{B592}}, \bibinfo{pages}{1}
  (\bibinfo{year}{2004}), \urlprefix\url{http://pdg.lbl.gov}.

\bibitem[{\citenamefont{Baym and Grinstein}(1977)}]{Baym:1977qb}
\bibinfo{author}{\bibfnamefont{G.}~\bibnamefont{Baym}} \bibnamefont{and}
  \bibinfo{author}{\bibfnamefont{G.}~\bibnamefont{Grinstein}},
  \bibinfo{journal}{Phys. Rev.} \textbf{\bibinfo{volume}{D15}},
  \bibinfo{pages}{2897} (\bibinfo{year}{1977}).

\bibitem[{\citenamefont{Bardeen and Moshe}(1986)}]{Bardeen:1986td}
\bibinfo{author}{\bibfnamefont{W.~A.} \bibnamefont{Bardeen}} \bibnamefont{and}
  \bibinfo{author}{\bibfnamefont{M.}~\bibnamefont{Moshe}},
  \bibinfo{journal}{Phys. Rev.} \textbf{\bibinfo{volume}{D34}},
  \bibinfo{pages}{1229} (\bibinfo{year}{1986}).

\bibitem[{\citenamefont{All{\'e}s et~al.}(1997)\citenamefont{All{\'e}s, D'Elia,
  and Di~Giacomo}}]{Alles:1996nm}
\bibinfo{author}{\bibfnamefont{B.}~\bibnamefont{All{\'e}s}},
  \bibinfo{author}{\bibfnamefont{M.}~\bibnamefont{D'Elia}}, \bibnamefont{and}
  \bibinfo{author}{\bibfnamefont{A.}~\bibnamefont{Di~Giacomo}},
  \bibinfo{journal}{Nucl. Phys.} \textbf{\bibinfo{volume}{B494}},
  \bibinfo{pages}{281} (\bibinfo{year}{1997}), \bibinfo{note}{erratum-ibid.
  {\bf B679}, 397 (2004)}, \eprint{hep-lat/9605013}.

\bibitem[{\citenamefont{All{\'e}s et~al.}(1999)\citenamefont{All{\'e}s, D'Elia,
  Di~Giacomo, and Stephenson}}]{Alles:1998mt}
\bibinfo{author}{\bibfnamefont{B.}~\bibnamefont{All{\'e}s}},
  \bibinfo{author}{\bibfnamefont{M.}~\bibnamefont{D'Elia}},
  \bibinfo{author}{\bibfnamefont{A.}~\bibnamefont{Di~Giacomo}},
  \bibnamefont{and} \bibinfo{author}{\bibfnamefont{P.~W.}
  \bibnamefont{Stephenson}}, \bibinfo{journal}{Nucl. Phys. Proc. Suppl.}
  \textbf{\bibinfo{volume}{73}}, \bibinfo{pages}{518} (\bibinfo{year}{1999}),
  \eprint{hep-lat/9808004}.

\bibitem[{\citenamefont{Chu and Schramm}(1995)}]{Chu:1994xz}
\bibinfo{author}{\bibfnamefont{M.~C.} \bibnamefont{Chu}} \bibnamefont{and}
  \bibinfo{author}{\bibfnamefont{S.}~\bibnamefont{Schramm}},
  \bibinfo{journal}{Phys. Rev.} \textbf{\bibinfo{volume}{D51}},
  \bibinfo{pages}{4580} (\bibinfo{year}{1995}), \eprint{nucl-th/9412016}.

\bibitem[{\citenamefont{Chu et~al.}(2000)\citenamefont{Chu, Ouellette, Schramm,
  and Seki}}]{Chu:1997tg}
\bibinfo{author}{\bibfnamefont{M.~C.} \bibnamefont{Chu}},
  \bibinfo{author}{\bibfnamefont{S.~M.} \bibnamefont{Ouellette}},
  \bibinfo{author}{\bibfnamefont{S.}~\bibnamefont{Schramm}}, \bibnamefont{and}
  \bibinfo{author}{\bibfnamefont{R.}~\bibnamefont{Seki}},
  \bibinfo{journal}{Phys. Rev.} \textbf{\bibinfo{volume}{D62}},
  \bibinfo{pages}{094508} (\bibinfo{year}{2000}), \eprint{hep-lat/9712023}.

\bibitem[{\citenamefont{Bernard et~al.}(1997)}]{Bernard:1996iz}
\bibinfo{author}{\bibfnamefont{C.~W.} \bibnamefont{Bernard}}
  \bibnamefont{et~al.}, \bibinfo{journal}{Phys. Rev. Lett.}
  \textbf{\bibinfo{volume}{78}}, \bibinfo{pages}{598} (\bibinfo{year}{1997}),
  \eprint{hep-lat/9611031}.

\bibitem[{\citenamefont{Chandrasekharan et~al.}(1999)}]{Chandrasekharan:1998yx}
\bibinfo{author}{\bibfnamefont{S.}~\bibnamefont{Chandrasekharan}}
  \bibnamefont{et~al.}, \bibinfo{journal}{Phys. Rev. Lett.}
  \textbf{\bibinfo{volume}{82}}, \bibinfo{pages}{2463} (\bibinfo{year}{1999}),
  \eprint{hep-lat/9807018}.

\bibitem[{\citenamefont{Gottlieb et~al.}(1997)}]{Gottlieb:1996ae}
\bibinfo{author}{\bibfnamefont{S.~A.} \bibnamefont{Gottlieb}}
  \bibnamefont{et~al.}, \bibinfo{journal}{Phys. Rev.}
  \textbf{\bibinfo{volume}{D55}}, \bibinfo{pages}{6852} (\bibinfo{year}{1997}),
  \eprint{hep-lat/9612020}.

\bibitem[{\citenamefont{Kogut et~al.}(1998)\citenamefont{Kogut, Lagae, and
  Sinclair}}]{Kogut:1998rh}
\bibinfo{author}{\bibfnamefont{J.~B.} \bibnamefont{Kogut}},
  \bibinfo{author}{\bibfnamefont{J.~F.} \bibnamefont{Lagae}}, \bibnamefont{and}
  \bibinfo{author}{\bibfnamefont{D.~K.} \bibnamefont{Sinclair}},
  \bibinfo{journal}{Phys. Rev.} \textbf{\bibinfo{volume}{D58}},
  \bibinfo{pages}{054504} (\bibinfo{year}{1998}), \eprint{hep-lat/9801020}.

\bibitem[{\citenamefont{Chen et~al.}(1998)}]{Chen:1998xw}
\bibinfo{author}{\bibfnamefont{P.}~\bibnamefont{Chen}} \bibnamefont{et~al.}
  (\bibinfo{year}{1998}), \eprint{hep-lat/9812011}.

\bibitem[{\citenamefont{Costa et~al.}(2005)\citenamefont{Costa, Ruivo,
  de~Sousa, and Kalinovsky}}]{Costa:2005cz}
\bibinfo{author}{\bibfnamefont{P.}~\bibnamefont{Costa}},
  \bibinfo{author}{\bibfnamefont{M.~C.} \bibnamefont{Ruivo}},
  \bibinfo{author}{\bibfnamefont{C.~A.} \bibnamefont{de~Sousa}},
  \bibnamefont{and} \bibinfo{author}{\bibfnamefont{Y.~L.}
  \bibnamefont{Kalinovsky}}, \bibinfo{journal}{Phys. Rev.}
  \textbf{\bibinfo{volume}{D71}}, \bibinfo{pages}{116002}
  (\bibinfo{year}{2005}), \eprint{hep-ph/0503258}.

\bibitem[{\citenamefont{Costa et~al.}(2004)\citenamefont{Costa, Ruivo,
  de~Sousa, and Kalinovsky}}]{Costa:2004db}
\bibinfo{author}{\bibfnamefont{P.}~\bibnamefont{Costa}},
  \bibinfo{author}{\bibfnamefont{M.~C.} \bibnamefont{Ruivo}},
  \bibinfo{author}{\bibfnamefont{C.~A.} \bibnamefont{de~Sousa}},
  \bibnamefont{and} \bibinfo{author}{\bibfnamefont{Y.~L.}
  \bibnamefont{Kalinovsky}}, \bibinfo{journal}{Phys. Rev.}
  \textbf{\bibinfo{volume}{D70}}, \bibinfo{pages}{116013}
  (\bibinfo{year}{2004}), \eprint{hep-ph/0408177}.

\bibitem[{\citenamefont{Michalski}(2006)}]{michalski06:_param_o_n}
\bibinfo{author}{\bibfnamefont{S.}~\bibnamefont{Michalski}},
  \bibinfo{journal}{in preparation}  (\bibinfo{year}{2006}).

\bibitem[{\citenamefont{Roh and Matsui}(1998)}]{Roh:1996ek}
\bibinfo{author}{\bibfnamefont{H.-S.} \bibnamefont{Roh}} \bibnamefont{and}
  \bibinfo{author}{\bibfnamefont{T.}~\bibnamefont{Matsui}},
  \bibinfo{journal}{Eur. Phys. J.} \textbf{\bibinfo{volume}{A1}},
  \bibinfo{pages}{205} (\bibinfo{year}{1998}), \eprint{nucl-th/9611050}.

\bibitem[{\citenamefont{M{\'o}csy et~al.}(2004)\citenamefont{M{\'o}csy,
  Mishustin, and Ellis}}]{Mocsy:2004ab}
\bibinfo{author}{\bibfnamefont{{\'A}.}~\bibnamefont{M{\'o}csy}},
  \bibinfo{author}{\bibfnamefont{I.~N.} \bibnamefont{Mishustin}},
  \bibnamefont{and} \bibinfo{author}{\bibfnamefont{P.~J.} \bibnamefont{Ellis}},
  \bibinfo{journal}{Phys. Rev.} \textbf{\bibinfo{volume}{C70}},
  \bibinfo{pages}{015204} (\bibinfo{year}{2004}), \eprint{nucl-th/0402070}.

\bibitem[{\citenamefont{Caldas et~al.}(2001)\citenamefont{Caldas, Mota, and
  Nemes}}]{Caldas:2000ic}
\bibinfo{author}{\bibfnamefont{H.~C.~G.} \bibnamefont{Caldas}},
  \bibinfo{author}{\bibfnamefont{A.~L.} \bibnamefont{Mota}}, \bibnamefont{and}
  \bibinfo{author}{\bibfnamefont{M.~C.} \bibnamefont{Nemes}},
  \bibinfo{journal}{Phys. Rev.} \textbf{\bibinfo{volume}{D63}},
  \bibinfo{pages}{056011} (\bibinfo{year}{2001}), \eprint{hep-ph/0005180}.

\end{thebibliography}

\end{document}